\documentclass[article,shortnames]{jss}
\usepackage{booktabs}      
\usepackage{dsfont}        

\author{Christian R\"{o}ver\\University Medical Center G\"{o}ttingen}
\title{Bayesian random-effects meta-analysis\\ using the \pkg{bayesmeta} \proglang{R} package}

\Plainauthor{Christian Roever} 
\Plaintitle{Bayesian random-effects meta-analysis using the bayesmeta R package} 
\Shorttitle{Meta-analysis using the bayesmeta package} 

\Abstract{

  The \emph{random-effects} or \emph{normal-normal hierarchical
    model} is commonly utilized in a wide range of meta-analysis
  applications.  A Bayesian approach to inference is very attractive
  in this context, especially when a meta-analysis is based only on
  few studies.  The \pkg{bayesmeta} \proglang{R}~package provides
  readily accessible tools to perform Bayesian meta-analyses and
  generate plots and summaries, without having to worry about
  computational details.  It allows for flexible prior specification
  and instant access to the resulting posterior distributions,
  including prediction and shrinkage estimation, and facilitating for
  example quick sensitivity checks.  The present paper introduces the
  underlying theory and showcases its usage.

}
\Plainkeywords{evidence synthesis, NNHM, between-study heterogeneity} 

   \Submitdate{2017-11-23}
   \Acceptdate{2018-09-02}

\Address{
  Christian R\"{o}ver\\
  Department of Medical Statistics\\
  University Medical Center G\"{o}ttingen\\
  Georg-August-Universit\"{a}t\\
  37073 G\"{o}ttingen, Germany\\
  E-mail: \email{christian.roever@med.uni-goettingen.de}\\
  URL: \url{http://www.gwdg.de/~croever}\\
  ORCID iD: \href{http://orcid.org/0000-0002-6911-698X}{0000-0002-6911-698X}
}

\providecommand{\realline}{\mathds{R}}
\providecommand{\prob}{\mathrm{P}}
\providecommand{\differential}{\mathrm{d}}
\providecommand{\expect}{\mathrm{E}}
\providecommand{\var}{\mathrm{Var}}
\providecommand{\normaldistn}{\mathrm{N}}

\providecommand{\condmu}{\widehat{\mu}}
\providecommand{\condsigma}{\widehat{\sigma}}

\providecommand{\priormu}{\mu_{\mathrm{p}}}
\providecommand{\priorsigma}{\sigma_{\mathrm{p}}}

\begin{document}
\section{Introduction}
\subsection{Meta-analysis}
  Evidence commonly comes in separate bits, and not necessarily from a
  single experiment.  In contemporary science, the careful conduct of
  systematic reviews of the available evidence from diverse data
  sources is an effective and ubiquitously practiced means of
  compiling relevant information.  In this context, meta-analyses
  allow for the formal, mathematical combination of information to
  merge data from individual investigations to a joint result.  Along
  with qualitative, often informal assessment and evaluation of the
  present evidence, meta-analytic methods have become a powerful tool
  to guide objective decision-making
  \citep{ChalmersHedgesCooper2002,LiberatiEtAl2009,HedgesOlkin,HartungKnappSinha,BorensteinEtAl}.
  Applications of meta-analytic methods span such diverse fields as
  agriculture, astronomy, biology, ecology, education, health
  research, medicine, psychology, and many more
  \citep{ChalmersHedgesCooper2002}.
  \looseness=-1    

  When empirical data from separate experiments are to be combined,
  one usually needs to be concerned about the straight one-to-one
  comparability of the provided results. There may be obvious or
  concealed sources of \emph{heterogeneity} between different
  studies, originating e.g.\ from differences in the selection and
  treatment of subjects, or in the exact definition of outcomes.
  Residual heterogeneity may be anticipated in the modeling stage and
  considered in the estimation process; a common approach is to
  include an additional variance component to account for
  between-study variability. On the technical side, the consideration
  of such a heterogeneity parameter leads to a \emph{random-effects}
  model rather than a \emph{fixed-effect} model
  \citep{HedgesOlkin,HartungKnappSinha,BorensteinEtAl}.  
%
%
  Inclusion of a non-zero heterogeneity will generally lead to more
  conservative results, as opposed to a ``na\"{i}ve'' merging of the
  given data without consideration of potentially heterogeneous data
  sources.

\subsection{The normal-normal hierarchical model (NNHM)}
  A wide range of problems may be approached using the
  \emph{normal-normal hierarchical model} (NNHM); this generic
  random-effects model is applicable when the estimates to be combined
  are given along with their uncertainties (standard errors) on a
  real-valued scale. Many problems are commonly solved this way, often
  after a transformation stage to re-formulate the problem on an
  appropriate scale. For example, binary data given in terms of
  contingency tables are routinely expressed in terms of logarithmic
  odds ratios (and associated standard errors), which are then readily
  processed via the NNHM\@.

  In the NNHM, measurements and standard errors are modeled via normal
  distributions, using means and their standard errors as sufficient
  statistics, while on a second hierarchy level the heterogeneity is
  modeled as an additive normal variance component as well. The model
  then has two parameters, the (real-valued) \emph{effect}, and the
  (positive) \emph{heterogeneity}. If the heterogeneity is zero,
  then the model reduces to the special case of a
  \emph{fixed-effect} model
  \citep{HedgesOlkin,HartungKnappSinha,HigginsGreen,BorensteinEtAl}.
  The model and terminology are described in detail in
  Section~\ref{sec:NNHM} below.
  The \pkg{bayesmeta} package is based on this simple yet ubiquitous
  form of the NNHM\@.

\subsection{Analysis within the NNHM framework}
\subsubsection{The Bayesian solution}
  The \pkg{bayesmeta} package implements a Bayesian approach to
  inference.  Bayesian modeling has previously been advocated and used
  in the meta-analysis context
  \citep{SmithSpiegelhalterThomas1995,SuttonAbrams2001,Spiegelhalter2004,SpiegelhalterEtAl,HigginsThompsonSpiegelhalter2009,LunnEtAl2013};
  the difference to the more common ``frequentist'' methods is that
  the problem is approached by expressing \emph{states of
    information} via probability distributions, where the
  consideration of new data then constitutes an update to a previous
  information state \citep{BDA3rd,Jaynes,SpiegelhalterEtAl1999}.  A
  Bayesian analysis allows (and in fact requires) the specification of
  \emph{prior information}, expressing the \emph{a~priori}
  knowledge, before data are taken into account. Technically this
  means the definition of a probability distribution, the
  \emph{prior distribution}, over the unknowns in the statistical
  model.  Once model and prior are specified, the results of a
  Bayesian analysis are uniquely determined; however, implementing the
  necessary computations to derive these in practice may still be
  tricky.

  While analysis results will of course depend on the prior
  setting, the range of reasonable specifications however is
  usually limited.  In the meta-analysis context, non-informative or
  weakly informative priors for the effect are readily defined, if
  required. For the between-study heterogeneity an informative
  specification is often appropriate, especially when only a small
  number of studies is involved.  Interestingly, the number of studies
  combined in the meta-analyses archived in the \emph{Cochrane Library} is
  reported by both \citet{DaveyEtAl2011} and
  \citet{KontopantelisSpringateReeves2013} with a median of~3 and a
  75\%~quantile of 6, so that in practice a majority of analyses 
  (including subgroup analyses and secondary outcomes)
  here is based on as few as 2--3 studies; such cases may not be as
  unusual as one might expect, at least in medical contexts.
  Typical meta-analysis sizes may vary across fields; for example, the
  data collected by \citet{vanErpEtAl2017} indicate a median number of
  12 and first and third quartiles of~5 and 33~studies, respectively,
  for meta-analyses published in the \emph{Psychological Bulletin}.
  Standard options for priors are available here, confining the prior
  probability within reasonable ranges \citep{SpiegelhalterEtAl}.
  Long-run properties of Bayesian methods have also been compared with
  common frequentist approaches by
  \citet{FriedeRoeverWandelNeuenschwander2017a,FriedeRoeverWandelNeuenschwander2017b},
  with a focus on the common case of very few studies.

  Bayesian methods commonly are computationally more demanding than
  other methods; usually these require the determination of
  high-dimensional integrals. In some (usually simpler) cases, the
  necessary integrals can be solved analytically, but it was mostly
  with the advent of modern computers and especially the development
  of Markov chain Monte Carlo (MCMC) methods that Bayesian analyses
  have become more generally tractable
  \citep{MetropolisUlam1949,MCMCinPractice}.  In the present case of
  random-effects meta-analysis within the NNHM, where only two unkown
  parameters are to be inferred, computations may be simplified by
  utilizing numerical integration or importance resampling
  \citep{TurnerEtAl2015}, both of which require relatively little
  manual tweaking in order to get them to work. It turns out that
  computations may be done partly analytically and partly numerically,
  offering another approach to simplify calculations via the
  \textsc{direct} algorithm \citep{RoeverFriede2017}. Utilizing this
  method, the \pkg{bayesmeta} package provides direct access to
  quasi-analytical posterior distributions without having to worry
  about setup, diagnosis or post-processing of MCMC algorithms. The
  present paper describes some of the methods along with the usage of
  the \pkg{bayesmeta} package.

\subsubsection{Other common approaches}
  A frequentist approach to inference is largely focused on long-run
  average operating characteristics of estimators.  In this framework,
  meta-analysis using the NNHM is most commonly done in two stages,
  where first the heterogeneity parameter is estimated, and then the
  effect estimate is derived based on the heterogenity estimate.  The
  choice of a heterogeneity estimator poses a problem on its own; a
  host of different heterogeneity estimators have been described, for
  a comprehensive summary of the most common ones see
  e.g.\ \citet{VeronikiEtAl2016}.  A common problem with such
  estimators of the heterogeneity variance component is that they
  frequently turn out as zero, effectively resulting in a fixed-effect
  model, which is usually seen as an undesirable feature.  Within this
  context, \citet{ChungEtAl2013a} proposed a penalized likelihood
  approach, utilizing a Gamma type ``prior'' penalty term in order to
  guarantee non-zero heterogeneity estimates.

  The treatment and estimation of heterogeneity in practice has been
  investigated e.g.\ by \citet{Pullenayegum2011},
  \citet{TurnerEtAl2012} and \citet{KontopantelisSpringateReeves2013}.
  When looking at large numbers of meta-analyses published by the
  Cochrane Collaboration, the majority (57\%) of heterogeneity
  estimates in fact turned out as zero \citep{TurnerEtAl2012}, while
  the numbers are higher for ``small'' meta-analyses, and lower for
  analyses involving many studies
  \citep{KontopantelisSpringateReeves2013}.  Meanwhile the choice of
  analysis method (fixed- or random-effects) also correlates with the
  number of studies involved, with larger numbers of studies
  increasing the chances of a random-effects model being employed
  \citep{KontopantelisSpringateReeves2013}.
  \citet{KontopantelisSpringateReeves2013} also compared the fraction
  of heterogeneity estimates resulting as zero in actual meta-analyses
  with that obtained from simulation, suggesting that heterogeneity is
  commonly underestimated or remains undetected.

  Once an estimate for the amount of heterogeneity has been arrived
  at, what is commonly done is to use this as a plug-in estimate and
  proceed to compute further tests and estimates \emph{conditioning
    on the heterogeneity estimate} as if its true value were known
  \citep{HedgesOlkin,HartungKnappSinha,BorensteinEtAl}.  Such a
  procedure would be warranted if the heterogeneity estimate was
  estimated with relatively great precision.  Notable exceptions here
  are the methods proposed by
  \citet{FollmannProschan1999,HartungKnapp2001a,HartungKnapp2001b} and
  \citet{SidikJonkman2002}, where the estimation uncertainty in
  heterogeneity is accounted for (on the technical side resulting in
  an inflated standard error and a heavier-tailed Student\mbox{-}$t$
  distribution to be utilized for deriving tests or confidence
  intervals), 
  or bootstrap methods \citep{NoortgateOnghena2005}
  and parameter estimation in the \emph{generalized inference} framework
  \citep{FriedrichKnapp2013}.

  \subsubsection{Bayesian and frequentist approaches in comparison}
  While the interpretation of results from a frequentist analysis,
  especially significance tests and confidence intervals, is commonly
  challenging and often misunderstood
  \citep{MoreyEtAl2016,HoekstraEtAl2014}, Bayesian results usually
  address the actual research question more directly and may be
  interpreted more intuitively
  \citep{Jaynes,SpiegelhalterEtAl,SzucsIoannidis2017,KruschkeLiddell2018}.
  On the one hand, frequentist confidence intervals aim to uniformly
  provide a pre-specified coverage probability 
  \emph{conditionally on any single point in parameter space}, 
  while Bayesian credible intervals account for the prior distribution
  and consequently provide proper coverage \emph{on average over the
    prior}
  (\citealt{Dawid1982}; see also Appendix~\ref{sec:calibration}, page~\pageref{sec:calibration}).  
  By their construction, they directly relate to the information on
  the parameters after considering the data at hand, which is not
  quite the intention behind classical confidence statements; even if a proper
  (``frequentist'') coverage probability is attained, this may still
  lead to rather counterintuitive conclusions in the face of actual data
  \citep{Jaynes1976,MoreyEtAl2016}.

  In some statistical applications, there is little difference between
  the results from frequentist and Bayesian analyses; often one may be
  considered a limiting or special case of the other, while
  interpretations remain somewhat different
  \citep{Bartholomew1965,Jaynes1976,Lindley1977,Severini1991,SpiegelhalterEtAl1999,BayarriBerger2004}.
  This is not necessarily the case in the present context, as
  meta-analyses are quite commonly based on few studies, so that
  certain large-sample asymptotics may not apply.  A common
  misconception, namely that a Bayesian analysis based on a uniform
  prior generally yielded identical results to a frequentist, purely
  likelihood-based analysis, is exposed as such here.  A crucial
  feature of meta-analysis problems is that one of the parameters, the
  heterogeneity, is confined to a bounded parameter space, which
  sometimes causes problems for frequentist methods
  \citep{Mandelkern2002}, partly because heterogeneity estimates
  commonly are not adequately characterized through a mere point
  estimate and an associated standard error. The common use of a
  plug-in estimate for the heterogeneity in frequentist procedures
  then turns out problematic, as such a strategy usually only makes
  sense when the estimated parameter is
  associated with relatively little uncertainty.
  Choice of a suitable heterogeneity estimator adds to the
  complication, as, despite their common aim, actual estimates may
  turn out quite differently, adding some degree of arbitrariness to
  the inference \citep{VeronikiEtAl2016}.
  Within a Bayesian context, these issues do not pose difficulties,
  and inference on some parameters while accounting for uncertainty in
  other nuisance parameters is straightforwardly solved through
  marginalization.
  This way, uncertainty in heterogeneity is readily accommodated, and
  since no asymptotic arguments need to be invoked, results are valid
  also for small sample sizes.  For such reasons Bayesian methods
  have been considered particularly well-suited for hierarchical
  models in general \citep{BrowneDraper2006,KruschkeLiddell2018}, and
  for meta-analysis problems in particular
  \citep{SmithSpiegelhalterThomas1995,SuttonAbrams2001}.  While
  Bayesian modeling necessitates the specification of a prior probability
  distribution over all parameters, the range of plausible
  formulations in a given context is usually limited.  Differences in
  results corresponding to different prior settings are quite natural,
  as effectively these correspond to differing answers to differently
  posed questions.

  Use of a coherent Bayesian framework also naturally facilitates
  advanced computations, in which the posterior from a previous
  analysis constitutes the prior for a subsequent analysis. This is
  useful for example in sequential meta-analyses
  \citep{SpenceEtAl2016}, in the design of future experiments
  \citep{SchmidliNeuenschwanderFriede2017}, or when utilizing
  historical data in the analysis of clinical trials
  \citep{WandelNeuenschwanderRoeverFriede2017}.

%
%

\subsubsection{Implementation}
  A number of software packages have been developed for frequentist
  inference within the NNHM framework, for example the \emph{Review
    Manager (RevMan)}, that is freely available from the Cochrane
  Collaboration \citep{RevMan5,HigginsGreen}, or, within \proglang{R},
  the \pkg{metafor} and \pkg{meta} packages
  \citep{Viechtbauer2010,Schwarzer2007,SchwarzerCarpenterRuecker}.

  Bayesian analyses are usually computationally more demanding, and
  quite generally these can be approached using MCMC methods
  \citep{MCMCinPractice}.  For example, meta-analysis along with the
  extension to meta-regression is implemented in the \pkg{bmeta}
  \proglang{R} package \citep{R:bmeta} by utilizing Gibbs sampling via
  JAGS \citep{Plummer2003,R:rjags}.  An MCMC approach offers great
  flexibility, and a number of model variations are also available,
  for example, a nonparametric generalization of the NNHM in the
  \pkg{bspmma} package \citep{R:bspmma}, a generalized approach based
  on model averaging in the \pkg{metaBMA} package \citep{R:metaBMA},
  or methods suitable for the special problem of meta-analysis of
  diagnostic studies in the \pkg{bamdit} and \pkg{metamisc} packages
  \citep{R:bamdit,R:metamisc}.

  In certain model constellations, it may be possible to derive exact
  posterior distributions, as for example implemented in the
  \pkg{mmeta} \proglang{R} package, which utilizes a parametric model
  for meta-analysis of count data that are provided in terms of
  contingency tables \citep{R:mmeta}. Otherwise, inference in a range
  of model classes may be approached via \emph{integrated nested
    Laplace approximations (INLA)}, as utilized e.g.\ in the
  \pkg{meta4diag} package for meta-analysis of diagnostic studies
  \citep{R:meta4diag}, or in the \pkg{nmaINLA} package for
  network-meta-analysis and \mbox{-}regression \citep{R:nmaINLA}.

  The \pkg{bayesmeta} package aims to provide easy access to a fully
  Bayesian analysis approach within the common NNHM framework. While
  the use of MCMC methods would be an option here, these usually
  require a certain amount of expertise and experience in set-up and
  convergence diagnostics. Also, inference based on MCMC output always
  contains a certain noise component due to the finite number of
  samples, which may sometimes constitute a nuisance. Use of the
  \pkg{bayesmeta} packages instantly provides accurate posterior
  summary figures analogous to output familiar from common
  (frequentist) meta-analysis output. Posterior distributions may be
  accessed in quasi-analytical form, and advanced methods, e.g.\ for
  prediction or shrinkage estimation, are also provided.  Computations
  are fast and reproducible, allowing for quick sensitivity checks and
  facilitating larger-scale simulations.
  Accuracy of the implementation (calibration) may be verified
  via simulation (see also Appendix~\ref{sec:calibration}).

%
%

\subsection{Outline}
  The remaining paper is mostly arranged in two major parts. In the
  following Section~\ref{sec:re-ma}, the underlying theory is
  introduced; first the common NNHM (random-effects) model and its
  notation are explained, and prior distributions for the two
  parameters are discussed. Then the resulting likelihood, marginal
  likelihood and posterior distributions are presented and some
  general points are introduced.

  In Section~\ref{sec:usage}, the actual usage of the \pkg{bayesmeta}
  package is demonstrated; an example data set is introduced, along
  which the steps of a Bayesian meta-analysis are shown.  The
  determination of summary statistics and plots, as well as possible
  variations in the analysis setup and the computation of posterior
  predictive $p$\mbox{-}values are presented.
  Section~\ref{sec:summary} then concludes with a summary.

\section{Random-effects meta-analysis}\label{sec:re-ma}
\subsection{The normal-normal hierarchical model}\label{sec:NNHM}
  The aim is to infer a quantity~$\mu$, the \emph{effect}, based on
  a number~$k$ of different measurements which are provided along with
  their corresponding uncertainties.  What is known are the empirical
  estimates~$y_i$ (of~$\mu$) that are associated with \emph{known}
  standard errors~$\sigma_i$; these constitute the ``input data''.
  The $i$th study's measurement~$y_i$ (where $i=1,\ldots,k$) is
  assumed to arise as exchangeable and normally distributed around the
  study's true parameter value~$\theta_i$:
  \begin{equation}\label{eqn:NNHM1}
    y_i|\theta_i,\sigma_i \; \sim \; \normaldistn(\theta_i, \sigma_i^2),
  \end{equation}
  where the variability is due to the sampling error, whose magnitude
  is given by the (known) standard error~$\sigma_i$.  All studies do
  not necessarily have identical true values~$\theta_i$; in order to
  accommodate potential between-study heterogeneity in the model, we
  assume that each study~$i$ measures a quantity~$\theta_i$ that
  differs from the overall mean~$\mu$ by another exchangeable,
  normally distributed offset with variance~$\tau^2 \geq 0$:
  \begin{equation}\label{eqn:NNHM2}
    \theta_i|\mu,\tau \; \sim \; \normaldistn(\mu,\tau^2).
  \end{equation}
  This second model stage implements the \emph{random effects}
  assumption.

  Especially when the study-specific parameters~$\theta_i$ are not of
  primary interest, the notation may be simplified by integrating out
  the ``intermediate'' $\theta_i$ terms and stating the model in its
  marginal form as
  \begin{eqnarray}\label{eqn:NNHM3}
    y_i|\mu,\tau,\sigma_i & \sim & \normaldistn(\mu,\,\sigma_i^2+\tau^2) \label{eqn:NNHM}
  \end{eqnarray}
  \citep{HedgesOlkin,HartungKnappSinha,BorensteinEtAl,BorensteinEtAl2010}.
  The two unknowns remaining to be inferred are the mean effect $\mu$
  and the heterogeneity~$\tau$, which is commonly considered a
  nuisance parameter. The studies' \emph{shrinkage estimates} of
  $\theta_i$ are however sometimes also of interest and may be
  inferred from the model as well.  In the special case of zero
  heterogeneity~($\tau=0$), the model simplifies to a fixed-effect
  model in which the study-specific means~$\theta_i$ are all identical
  ($\theta_1=\ldots=\theta_k=\mu$).

  Such two-stage hierarchical models of an overall mean ($\mu$) and
  study-specific parameters ($\theta_i$) with a random effect for each
  study are commonly utilized in meta-analysis applications. The
  simple case of normally distributed error terms at both stages is
  often convenient and easily tractable, and it also constitutes a
  good approximation in many cases.  So, while the effect here is
  treated as a continuous parameter, the model is quite commonly
  utilized to also process different types of data (e.g.\ logarithmic
  odds ratios from dichotomous data, etc.) after transformation to a
  real-valued effect scale
  \citep{HedgesOlkin,HartungKnappSinha,BorensteinEtAl,Viechtbauer2010,HigginsGreen}.

\subsection{Prior distributions}\label{sec:prior}
\subsubsection{General}\label{sec:priorGeneral}
  Among the two unknowns, the effect $\mu$ is commonly of primary
  interest, while the heterogeneity~$\tau$ usually is considered a
  nuisance parameter.  In order to infer the parameters, we need to
  specify our prior information about~$\mu$ and~$\tau$ in terms of
  their joint prior probability density function~$p(\mu, \tau)$.  What
  exactly constitutes a reasonable prior distribution always depends
  on the given context \citep{BDA3rd,SpiegelhalterEtAl,Jaynes}.  For
  computational convenience, in the following we assume that we can
  factor the prior density into independent marginals: $p(\mu, \tau) =
  p(\mu) \times p(\tau)$. While this may not seem unreasonable,
  depending on the context, one may also argue in favour of a
  dependent prior specification
  \citep[e.g.,][]{Senn2007,Pullenayegum2011}.  In the following, we
  aim to provide a comprehensive overview of popular or sensible
  options.  We will discuss proper as well as improper priors; when
  using improper priors, the usual care must be taken, as the
  resulting posterior then may or may not be a proper probability
  distribution \citep{BDA3rd}.  The discussed heterogeneity priors are
  also summarized in Table~\ref{tab:priors} below.

\subsubsection[The effect parameter]{The effect parameter $\mu$}\label{sec:effectPrior}
  An obvious choice of a non-informative prior for the effect~$\mu$,
  being a location parameter, is an improper uniform distribution over
  the real line \citep{BDA3rd,SpiegelhalterEtAl,Jaynes}. A normal
  prior (with mean~$\priormu$ and variance~$\priorsigma^2$) is a
  natural choice as an informative prior for the effect~$\mu$, and
  these two are also the cases we will restrict ourselves to for
  computational convenience and feasibility in the following.  The
  normal prior here constitutes the conditionally conjugate prior
  distribution for the effect (see also Section~\ref{sec:conditional}
  below).  The uninformative uniform prior would also result as the
  limiting case for increasing prior uncertainty
  ($\priorsigma\rightarrow\infty$).

  A way to guide the choice of a vague prior is by consideration of
  \emph{unit information priors} \citep{KassWasserman1995}. The idea
  here is to specify the prior such that its information content
  (variance) is in some way, possibly somewhat heuristically,
  equivalent to a single observational unit.  For example, if the
  endpoint is a logarithmic odds ratio (log-OR), a neutral unit
  information prior may be given by a normal prior with zero mean
  (centered around an odds ratio of~1, i.e., ``no effect'') and a
  standard deviation of~$\priorsigma=4$.  For a derivation, see also
  Appendix~\ref{sec:unitInfoApp}.

\subsubsection[The heterogeneity parameter: proper, informative priors]{The heterogeneity parameter $\tau$: proper, informative priors}
  Especially since in the meta-analysis context one is commonly
  dealing with very small numbers of studies~$k$, where not much
  information on between-study heterogeneity may be expected to be
  gained from the data, it may be worth while considering the use of
  informative priors. Depending on the exact context, there often is
  some information on what values for the heterogeneity are more
  plausible and which ones are less so, and making use of the present
  information may make a difference in the end.  For example, if the
  meta-analysis is based on logarithmic odds ratios, it will usually
  make sense to assume that heterogeneity is unlikely to exceed, say,
  $\tau\!=\!\log(10)\!\approx\!2.3$, which would correspond to roughly
  an expected factor~10 difference in effects (odds ratios) between
  trials due to heterogeneity.  An extensive discussion of such cases
  is provided in \citet[][Sec.~5.7]{SpiegelhalterEtAl}. Values for
  $\tau$ between 0.1 and 0.5 here are considered ``reasonable'',
  values between 0.5 and 1.0 are ``fairly high'' and values beyond 1.0
  are ``fairly extreme''.  An analogous reasoning would apply for
  similarly defined outcomes, for example, logarithmic relative risks,
  logarithmic hazard ratios, or logarithmic variances
  \citep{SchmidliNeuenschwanderFriede2017}.  Consideration of the
  magnitude of unit information variances (see previous paragraph) may
  also be helpful in this context, as variability (heterogeneity)
  between studies will usually be expected to be substantially below
  the variability between individuals.  Along these lines, it is often
  useful to also consider the implications of prior specifications in
  terms of the corresponding \emph{prior predictive distributions};
  see also Section~\ref{sec:priorVariations} below.  The impact of
  variations of how exactly prior information is implemented in the
  model may eventually also be checked via sensitivity analyses.

  A sensible informative choice for $p(\tau)$ may be the maximum
  entropy prior for a pre-specified prior expectation~$\expect[\tau]$,
  the exponential distribution with rate
  $\lambda=\frac{1}{\expect[\tau]}$ \citep{Jaynes1968,Jaynes,Gregory}.
  Log-normal or half-normal prior distributions, e.g.\ with
  pre-specified quantiles, may also be useful alternatives.  For
  example, for log-OR (or similar) endpoints, the routine use of
  half-normal distributions with scale~$0.5$ or~$1.0$ has been
  suggested by
  \citet{FriedeRoeverWandelNeuenschwander2017a,FriedeRoeverWandelNeuenschwander2017b}
  and was shown to work well in simulations.  In order to gain
  robustness, one may also consider mixture distributions as
  informative priors, for example half-Student\mbox{-}$t$,
  half-Cauchy, or Lomax distributions, which may be considered
  heavy-tailed variants of half-normal or exponential distributions
  \citep{JohnsonKotzBalakrishnan}.  Use of a heavy-tailed prior
  distribution will allow for discounting of the prior in favour of
  the data in case the data appear to be in conflict with prior
  expectations \citep{OHaganPericchi2012,SchmidliEtAl2014}.  The use
  of weakly informative half-Student\mbox{-}$t$ or half-Cauchy priors
  may also be motivated via theoretical arguments, as these can be
  shown to also exhibit favourable frequentist properties
  \citep{Gelman2006,PolsonScott2012}.

  Although an inverse-Gamma distribution for an informative prior may
  seem to be an obvious choice, use of this distribution is generally
  \emph{not} recommended \citep{Gelman2006,PolsonScott2012}.  More
  on informative (as well as uninformative) priors may be found in
  \citet{SpiegelhalterEtAl}, \citet{Gelman2006} and
  \citet{PolsonScott2012}.  Some empirical evidence to consider for
  informative priors for certain types of endpoints may be found
  e.g.\ in \citet{Pullenayegum2011}, \citet{TurnerEtAl2012},
  \citet{KontopantelisSpringateReeves2013} and
  \citet{vanErpEtAl2017}. In particular, \citet{RhodesEtAl2015} and
  \citet{TurnerEtAl2015} derived empirical priors based on data from
  the \emph{Cochrane database of systematic reviews}; prior
  information here is expressed in terms of log-normal or
  log-Student\mbox{-}$t$ distributions.

\subsubsection[The heterogeneity parameter: proper, `non-informative' priors]{The heterogeneity parameter $\tau$: proper, `non-informative' priors}
\label{sec:ProperNoninf}
  Some ``non-informative'' proper priors have been proposed that are
  scale-in\-vari\-ant in the sense that (like the Jeffreys prior
  discussed below as well) they depend only on the standard
  errors~$\sigma_i$. A re-ex\-press\-ion of the estimation problem on
  a different measurement scale would entail a proportional re-scaling
  of standard errors and so inference effectively remains
  unaffected. Such priors are discussed e.g.\ by
  \citet[Sec.~5.7.3]{SpiegelhalterEtAl} and \citet{BergerDeely1988}.
  Priors like these, however, are somewhat problematic from a logical
  perspective, as these imply that the prior information on the
  heterogeneity depended on the accuracy of the individual studies'
  estimates \citep{Senn2007}.

  The following two priors both depend on the harmonic mean~$s_0^2$ of
  squared standard errors, i.e.,
  \begin{equation} \label{eqn:s02}
    s_0^2 \;=\; \frac{k}{\sum_{i=1}^k \sigma_i^{-2}}.
  \end{equation}
  The \emph{uniform shrinkage prior} results from considering the
  ``average shrinkage'' $S(\tau)=\frac{s_0^2}{s_0^2+\tau^2}$; placing
  a uniform prior on~$S(\tau)$ results in a prior density
  \begin{equation} \label{eqn:UnifShrink}
    p(\tau) \;=\; \frac{2\tau s_0^2}{\bigl(s_0^2+\tau^2\bigr)^2}
  \end{equation}
  for the heterogeneity, which has a median of $s_0$.  For a detailed
  discussion see e.g.\ \citet{SpiegelhalterEtAl} or
  \citet{Daniels1999}.  A uniform prior in $S(\tau)$ is equivalent to
  a uniform prior in $1\!-\!S(\tau)=\frac{\tau^2}{s_0^2+\tau^2}$
  \citep{SpiegelhalterEtAl}, which is an expression very similar to
  the $I^2$~measure of heterogeneity due to
  \citet{HigginsThompson2002}.  Substiting the harmonic mean~$s_0^2$
  for their average ($\hat{s}^2$) in the prior
  density~(\ref{eqn:UnifShrink}) hence yields a uniform prior
  in~$I^2$.
  
  The \emph{DuMouchel prior} has a similar form and is defined
  through
  \begin{equation} \label{eqn:DuMouchel}
    p(\tau) \;=\; \frac{s_0}{(s_0+\tau)^2}.
  \end{equation}
  This implies a log-logistic distribution for the
  heterogeneity~$\tau$ that has its mode at $\tau=0$ and its median at
  $\tau=s_0$ \citep{SpiegelhalterEtAl,DuMouchelNormand2000}.

  A \emph{conventional prior} as a proper variation of the Jeffreys
  prior (see also the closely related variant
  in~(\ref{eqn:BergerDeely}) below) was given by
  \citet{BergerDeely1988} as
  \begin{equation}\label{eqn:conventional}
    p(\tau) \;\propto\; \prod_{i=1}^k \Biggl(\frac{\tau}{\bigl(\sigma_i^2+\tau^2\bigr)^{3/2}}\Biggr)^{1/k}.
  \end{equation}
  This prior is in particular intended as a non-informative but proper
  choice for testing or model selection purposes
  \citep{BergerDeely1988,BergerPericchi2001}.
  \vspace{2ex}    

\subsubsection[The heterogeneity parameter: improper priors]{The heterogeneity parameter $\tau$: improper priors}
\paragraph{Uninformative priors}
  It is not so obvious what exactly would qualify a prior for~$\tau$
  as ``uninformative''.  One might argue that an uninformative prior
  should have a probability density function that is monotonically
  decreasing in~$\tau$; another question would be whether the
  density's intercept $p(\tau=0)$ should be positive or finite, or
  what the density's derivative near zero should be.  In general, the
  uninformative prior for a scale parameter in a simple normal model
  is commonly taken to be uniform in $\log(\tau)$ (and $\log(\tau^2)$)
  with density $p(\tau)\propto\frac{1}{\tau}$
  \citep{Jeffreys1946,BDA3rd}, however, this ``log-uniform'' prior
  will not lead to proper, integrable posteriors in the present
  context \citep{Gelman2006}.  Another reasonable choice may be the
  improper uniform prior on the positive real line, but care must be
  taken here as usual, as the posterior may end up improper as well;
  this will not result in a proper posterior when there are only one
  or two estimates available (i.e., when $k\leq 2$) and an (improper)
  uniform effect prior is used \citep{Gelman2006}.  The uniform prior
  may be considered a conservative choice in a particular sense, as
  shown below (Appendix~\ref{sec:conservative}), but on the other hand
  it may also be considered overly conservative, as it tends to attach
  a lot of weight to potentially unreasonably large heterogeneity
  values. \citet{Gelman2006} generally recommends a uniform
  heterogeneity prior as an uninformative default, \emph{unless} the
  number of studies~$k$ is small, or an informative prior is desired
  or for other reasons.

  One may also argue via certain requirements that an uninformative
  prior should meet \citep{Jaynes1968,Jaynes}. For example, it may be
  reasonable to demand invariance with respect to re-scaling of $\tau$
  for the prior density~$p(\tau)$, leading to a constraint of the form
  \begin{equation} \label{eqn:uninfoCondition}
     \textstyle \frac{1}{s} \; p\bigl(\frac{\tau}{s}\bigr) \;=\; f(s) \; p(\tau)
  \end{equation}
  for any scaling factor $s>0$ and some positive-valued function
  $f(s)$ (i.e., re-scaling should not affect the density's shape).
  This requirement obviously restricts the range of priors to those
  with monotonic density functions.  It leads to a family of improper
  prior distributions with densities
  \begin{equation}\label{eqn:PowerAPrior}
    p(\tau) \;\propto\; \tau^a
  \end{equation}
  for $a\in\realline$.  This family includes (for~$a\!=\!-1$) the
  common log-uniform prior for a scale parameter mentioned above, or
  (for~$a\!=\!0$) the uniform prior.  But this class also includes
  further interesting cases, like, for $-1<a<0$, a compromise between
  the above two uniform and log-uniform priors that is (locally)
  integrable over any interval $[0,u]$ with $0 < u < \infty $ while
  also being shorter-tailed on the right than the improper uniform
  prior. An obvious example is (for $a=-0.5$) the prior with
  monotonically decreasing density function
  \begin{equation}\label{eqn:sqrtPrior}
    p(\tau) \;\propto\; \textstyle \frac{1}{\sqrt{\tau}}
  \end{equation}
  which corresponds to a uniform prior in $\sqrt{\tau}$.  This prior
  has the unusual property that the prior density, and with that the
  posterior as well, exhibits a pole (i.e., approaches infinity) at
  the origin.  A value of $a\!=\!1$ would lead to a uniform prior
  in~$\tau^2$, with an even higher preference for large heterogeneity
  values, which requires at least $k\!\geq\!4$ studies for a proper
  posterior; this prior is generally not recommended
  \citep{Gelman2006}.

\paragraph{The Jeffreys prior}
  The non-informative Jeffreys prior \citep{BDA3rd,Jeffreys1946} for
  this problem results from the form of the likelihood (see
  equation~(\ref{eqn:NNHM}) or (\ref{eqn:jointLogLikeli}) below), or
  more specifically, the associated expected Fisher
  information~$J(\mu,\tau)$; its probability density function is given
  by $ p(\mu, \tau) \propto \sqrt{\det\bigl(J(\mu,\tau)\bigr)}$.  This
  general form of Jeffreys' prior however is generally not recommended
  when the set of parameters includes a location parameter as in the
  present case; see e.g.\ \citet{Jeffreys1946},
  \citet[Sec.~III.3.10]{Jeffreys}, \citet[Sec.~3.3.3]{Berger} and
  \citet[Sec.~2.2]{KassWasserman1996}.  Instead, location parameters
  are commonly treated as fixed and are conditioned upon
  \citep{Berger,KassWasserman1996}. In the present case (since $\mu$
  and $\tau$ are orthogonal in the sense that the Fisher information
  matrix' off-diagonal elements are zero), this leads to Tibshirani's
  non-informative prior
  \citep[Sec.~3.7]{Tibshirani1989,KassWasserman1996}, a variation of
  the general Jeffreys prior, which is of the form
  \begin{equation}\label{eqn:Jeffreys}
    p(\tau) \;\propto\;
    \sqrt{\sum_{i=1}^k \Bigl(\frac{\tau}{\sigma_i^2+\tau^2}\Bigr)^2}
    \mbox{.}
  \end{equation}
  In the following, we will consider this variant as the
  \emph{Jeffreys prior} for the NNHM\@.  This prior also constitutes
  the \emph{Berger-Bernardo reference prior} for the present problem
  \citep{BodnarLinkElster2016,BodnarEtAl2017}.  The prior is improper,
  as the right tail asymptotically behaves like
  $p(\tau)\propto\frac{1}{\tau}$, but it is locally integrable in the
  left tail with $p(0)=0$.  The resuling posterior is proper as long
  as $k\geq 2$ \citep{BodnarEtAl2017}.

  In case of constant standard errors $\sigma_i\!=\!\sigma$, the
  prior's mode is at $\tau\!=\!\sigma$.  Otherwise the mode tends to
  be near the smallest~$\sigma_i$, but the prior may also be
  multimodal. The Jeffreys prior's dependence on the standard
  errors~$\sigma_i$ implies that the prior information varies with the
  precision of the underlying data~$y_i$. With greater precision,
  lower heterogeneity values are considered plausible. On the other
  hand, the prior is invariant to the overall scale of the problem (as
  it scales with the standard errors~$\sigma_i$) like the proper
  non-informative priors mentioned above.

  Another variation of the Jeffreys prior was given by
  \citet{BergerDeely1988} and is defined as
  \begin{equation} \label{eqn:BergerDeely}
    p(\tau) \;\propto\; \prod_{i=1}^k \biggl(\frac{\tau}{\sigma_i^2+\tau^2}\biggr)^{1/k}.
  \end{equation}
  This prior is also improper, and it equals the Jeffreys prior in
  case all standard errors $\sigma_i$ are identical.

\subsubsection{Choice of a prior}
  The selection of a prior for the effect~$\mu$ is relatively
  straightforward. The normal prior's variance allows to vary the
  width from narrow/informative to wide/uninformative; the improper
  uniform prior as a limiting case is also available, and this may be
  the obvious default choice in many cases.  Consideration of the unit
  information prior's width may also help judging the amount of
  information conveyed by a given informative prior.

  The heterogeneity priors discussed above may roughly be categorized
  in four classes, as shown in Table~\ref{tab:priors}. First of all,
  one needs to decide whether a proper prior is desired or
  required. Arguments in favour of a proper prior may include the need
  for finite marginal likelihoods and Bayes factors in model selection
  problems, general preference, or a small number ($k$) of studies.

  {
  \begin{table}[h]
  \begin{center}
    \caption{\label{tab:priors} The heterogeneity priors discussed in
      Section~\ref{sec:prior} may roughly be divided into 4~classes;
      some of their properties are summarized
      below.} \vspace{1ex}\small
    \begin{tabular}{lllll}
      \toprule
                        & \multicolumn{2}{c}{proper} & \multicolumn{2}{c}{improper} \\ \addlinespace[-0.3ex]
      \cmidrule(lr){2-3} \cmidrule(lr){4-5}
       & informative & non-informative & non-informative & scale-invariant \\
      \midrule
      examples                  & \begin{minipage}[t]{0.17\textwidth}\raggedright
                                    half-normal, half-Student\mbox{-}$t$,
                                    half-Cauchy, log-normal, exponential, \ldots
                                  \end{minipage}
                                & \begin{minipage}[t]{0.17\textwidth}\raggedright
                                    uniform \mbox{$\quad$}shrinkage (\ref{eqn:UnifShrink}), DuMouchel (\ref{eqn:DuMouchel}), conventional (\ref{eqn:conventional})
                                  \end{minipage}
                                & \begin{minipage}[t]{0.17\textwidth}\raggedright
                                    Jeffreys (\ref{eqn:Jeffreys}), \mbox{Berger-Deely (\ref{eqn:BergerDeely})}
                                  \end{minipage}
                                & \begin{minipage}[t]{0.17\textwidth}\raggedright
                                    uniform in~$\tau$, uniform in~$\sqrt{\tau}$, {\ldots} (\ref{eqn:PowerAPrior})
                                  \end{minipage}\\
      dependent on $\sigma_i$? & no & yes & yes & no \\
      scale-invariant?          & no & scales with $\sigma_i$ & scales with $\sigma_i$ & yes \\
      $k$ restrictions?         & --- & $k\geq 1$  & $k\geq 2^\ast$ & $k\geq 3^\ast$ \\
      \bottomrule
      \multicolumn{5}{r}{$^\ast${\footnotesize}(less if combined with a proper effect prior)}
    \end{tabular}
  \end{center}
  \end{table}}

  Among the proper priors one then has the choice between informative
  distributions, and priors that are supposed to be non-informative,
  which however depend on the involved studies' standard
  errors~$\sigma_i$.  The improper priors discussed here are all
  uninformative in one or another sense; the Jeffreys and Berger-Deely
  priors also depend on the~$\sigma_i$, they require at least
  \mbox{$k=2$}~available studies, the uniform prior is independent of
  the~$\sigma_i$ and requires at least $3$~studies.

\begin{figure}[t]
    \begin{center}
      \includegraphics[width=0.7\linewidth]{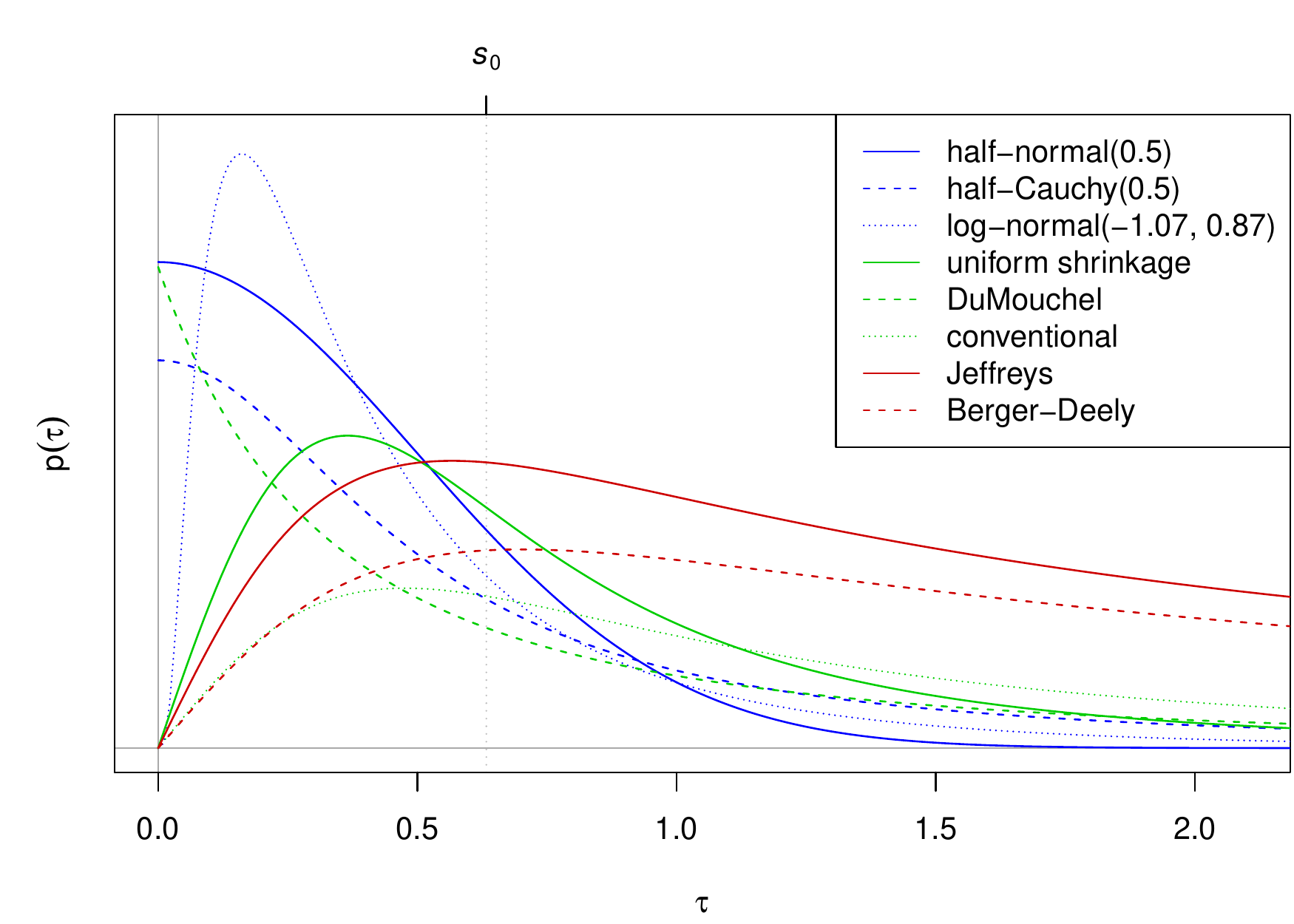}
      \caption{\label{fig:priors}A selection of prior distributions
        for the example data discussed in Section~\ref{sec:usage}
        below.  Half-normal and half-Cauchy parameters are reasonable
        choices for log-OR endpoints.  The log-normal parameters are
        chosen according to \citet{TurnerEtAl2015}.  The uniform
        shrinkage and DuMouchel priors are scaled relative to the
        harmonic mean of squared standard errors~$s_0^2$.  The Jeffreys and
        Berger-Deely priors are improper, so their densities do not
        integrate to a finite value.}
    \end{center}
  \end{figure}

  Some prior densities are illustrated in Figure~\ref{fig:priors}. As
  the choice of a sensible informative prior depends on the context,
  and some other priors depend on the $\sigma_i$~values, the priors
  shown here correspond to the example discussed in
  Section~\ref{sec:usage} below. The proper informative half-normal
  and half-Cauchy priors with scale~0.5 are reasonable choices for
  log-ORs and similar endpoints. The log-normal prior's parameters are
  recommended for the type of investigation based on the analysis
  by~\citet{TurnerEtAl2015}. The proper uniform shrinkage, DuMouchel
  and conventional priors depend on the involved studies' standard
  errors~$\sigma_i$. The improper Jeffreys and Berger-Deely prior
  densities do not integrate to a finite value, so their overall
  scaling is somewhat arbitrary here.

  \citet{Gelman2006} generally recommends the improper uniform
  heterogeneity prior, \emph{unless} the number of studies~$k$ is
  small, or an informative prior is desired or for other reasons. In
  those cases, an informative prior from the half-Student\mbox{-}$t$
  family is recommended, which includes half-Cauchy and half-normal
  priors as special or limiting cases. Use of the half-Cauchy family
  is further supported by \citet{PolsonScott2012} based also on
  classical frequentist properties.  If, for example, the endpoint is
  a log-OR, then, using the categorization by
  \citet[][Sec.~5.7]{SpiegelhalterEtAl}, a half-normal prior with
  scale 0.5 may confine heterogeneity mostly to ``reasonable'' to
  ``fairly high'' values and leave about 5\% probability for ``fairly
  extreme'' heterogeneity. A larger scale parameter or a
  heavier-tailed distribution may then serve as a more conservative or
  more robust reference for a sensitivity check
  \citep{FriedeRoeverWandelNeuenschwander2017a,FriedeRoeverWandelNeuenschwander2017b}.
  The Jeffreys prior constitutes another default choice of an
  uninformative prior; as the \emph{Berger-Bernardo reference
    prior} it represents the least informative prior in a certain
  sense \citep{BodnarEtAl2017}, and it will yield a proper posterior
  as long as at least 2~studies are available.
  \vspace{1ex}    

\subsection{Likelihood}\label{sec:likelihood}
  The form of the likelihood follows from the assumptions introduced
  in Section~\ref{sec:NNHM}. The NNHM is essentially a simple normal
  model with unknown mean and an unknown variance component; the
  resulting likelihood function is given by
  \begin{equation}
    p(\vec{y}|\mu,\tau,\vec{\sigma}) 
    \;=\; 
    (2\pi)^{-\frac{k}{2}} \times
    \prod_{i=1}^k\frac{1}{\sqrt{\sigma_i^2+\tau^2}}
    \exp\biggl(-\frac{1}{2}\frac{(y_i-\mu)^2}{\sigma_i^2+\tau^2}\biggr)\mbox{,}
    \label{eqn:jointLikeli}
  \end{equation}
  where $\vec{y}$ and $\vec{\sigma}$ denote the vectors of $k$~effect
  measures~$y_i$ and their standard errors~$\sigma_i$. For any
  practical application it is often more useful to consider the
  logarithmic likelihood, i.e.,
  \begin{equation}
    \log\bigl(p(\vec{y}|\mu,\tau,\vec{\sigma})\bigr)
    \;=\; 
    -{\textstyle \frac{k}{2}}\log(2\pi)
    -{\textstyle \frac{1}{2}}\sum_{i=1}^k\biggl( \log(\sigma_i^2+\tau^2)
                      + \frac{(y_i-\mu)^2}{\sigma_i^2+\tau^2} \biggr)\mbox{.}
    \label{eqn:jointLogLikeli}
  \end{equation}

\subsection{Marginal likelihood}\label{sec:marginal}
\subsubsection{Marginalization}
  In order to do inference within a Bayesian framework, it is usually
  necessary to compute integrals involving the posterior distribution
  \citep{BDA3rd}. For example, in a multi-parameter model, one may be
  interested in the \emph{marginal posterior distribution} or in the
  \emph{posterior expectation} of a certain parameter, both of which
  result as integrals. Key to the \pkg{bayesmeta} implementation is
  the partly analytical and partly numerical integration over
  parameter space. In the following, we will derive the marginal
  posterior distribution of the heterogeneity parameter via the
  marginal likelihood, and we will later see how marginal and
  conditional distributions may be utilized to evaluate the required
  integrals.  The likelihood is initially a function of both
  parameters ($\mu$ and $\tau$), and the marginal likelihood of the
  heterogeneity~$\tau$ results from integration over the effect~$\mu$,
  using its prior distribution, which we specified to be either
  uniform or normal.
  \vspace{3ex}    

\subsubsection{Uniform prior}
  Using the improper uniform prior for the effect~$\mu$
  ($p(\mu)\propto 1$), we can derive the marginal likelihood,
  marginalized over~$\mu$,
\begin{eqnarray}
  p(\vec{y}|\tau,\vec{\sigma}) 
  &=& 
  \int p(\vec{y}|\mu,\tau,\vec{\sigma}) \, p(\mu) \, \differential \mu \mbox{.}
\end{eqnarray}
  For the NNHM, the integral turns out as
\begin{eqnarray}
  p(\vec{y}|\tau,\vec{\sigma}) 
  &=&
  \bigl(2\pi\bigr)^{-\frac{k-1}{2}} \times
  \prod_{i=1}^k\frac{1}{\sqrt{\sigma_i^2+\tau^2}} \times
  \exp\biggl(-\frac{1}{2}\frac{\bigl(y_i-\condmu(\tau)\bigr)^2}{\sigma_i^2+\tau^2}\biggr)
  \times
  \frac{1}{\sqrt{\sum_{i=1}^k \frac{1}{\sigma_i^2+\tau^2}}}\mbox{,}
  \label{eqn:marginal1b}
\end{eqnarray}
  where $\condmu(\tau)$ is the conditional
  posterior mean of $\mu$ for a given
  heterogeneity $\tau$. Conditional mean and standard deviation are given by
\begin{equation}\label{eqn:conditionalmoments1}
  \condmu(\tau)
  \;=\;
  \expect[\mu|\tau,\vec{y},\vec{\sigma}]
  \;=\;
  \frac{\sum_{i=1}^k\frac{y_i}{\sigma_i^2+\tau^2}}{\sum_{i=1}^k\frac{1}{\sigma_i^2+\tau^2}}
  \qquad \mbox{and} \qquad
  \condsigma(\tau)
  \;=\;
  \sqrt{\var(\mu|\tau,\vec{y},\vec{\sigma})}
  \;=\;
  \sqrt{\frac{1}{\sum_{i=1}^k\frac{1}{\sigma_i^2+\tau^2}}} 
  \mbox{.} 
\end{equation}
  A derivation is provided in Appendix~\ref{sec:marginalDerivation};
  the standard deviation~$\condsigma(\tau)$ will become relevant later on.
  On the logarithmic scale the marginal likelihood then is:
\begin{eqnarray}
  &&\log\bigl(p(\vec{y}|\tau,\vec{\sigma})\bigr) \nonumber \\
  &=&
  -\frac{1}{2}\Biggl( (k\!-\!1)\log(2\pi) +
    \sum_{i=1}^k\biggl(\log\bigl(\sigma_i^2\!+\!\tau^2\bigr)
                      +\frac{\bigl(y_i - \condmu(\tau)\bigr)^2}{\sigma_i^2+\tau^2}\biggr)
    + \log\biggl(\sum_{i=1}^k\frac{1}{\sigma_i^2\!+\!\tau^2}\biggr) \Biggr)\mbox{.}
  \label{eqn:marginal1c}
\end{eqnarray}
  \vspace{3ex}    

\subsubsection{Conjugate normal prior}
  The normal effect prior here is the \emph{conditionally} conjugate
  prior distribution, since the resulting conditional posterior (for a
  given $\tau$~value) again is of a normal form.  Calculations for the
  (proper) normal prior for the effect~$\mu$ work similarly to the
  previous derivation.  Assume the prior for~$\mu$ is normal with
  mean~$\priormu$ and variance~$\priorsigma^2$, i.e., it is defined
  through the probability density function
  $p(\mu)=\frac{1}{\sqrt{2\pi}\,
    \priorsigma}\exp\Bigl(-\frac{1}{2}\frac{(\mu-\priormu)^2}{\priorsigma^2}\Bigr)$.
  The necessary integral for the marginal likelihood then results as
\begin{eqnarray}
  p(\vec{y}|\tau,\vec{\sigma}) 
  &=& 
  \int p(\vec{y}|\mu,\tau,\vec{\sigma}) \, p(\mu) \, \differential \mu
  \\
  &=&
  \bigl(2\pi\bigr)^{-\frac{k+1}{2}} 
  \times
  \frac{1}{\sqrt{\priorsigma^2}}
  \times
  \prod_{i=1}^k\frac{1}{\sqrt{\sigma_i^2+\tau^2}} 
  \nonumber \\
  &&\times
  \int \exp\biggl(-\frac{1}{2}\Bigl[\frac{(\mu-\priormu)^2}{\priorsigma^2} + \sum_{i=1}^k\frac{(y_i-\mu)^2}{\sigma_i^2+\tau^2}\Bigr]\biggr) \, \differential \mu. \label{eqn:marginal2a}
\end{eqnarray}
  One can see that the prior parameters ($\priormu$ and $\priorsigma$)
  enter in a similar manner as the data points ($y_i$ and
  $\sigma_i$). In analogy to the previous derivation, define the
  conditional posterior mean and standard deviation
\begin{equation}\label{eqn:conditionalmoments2}
  \condmu(\tau)
  \;=\;
  \frac{\frac{\priormu}{\priorsigma^2}+\sum_{i=1}^k\frac{y_i}{\sigma_i^2+\tau^2}}{\frac{1}{\priorsigma^2}+\sum_{i=1}^k\frac{1}{\sigma_i^2+\tau^2}} 
  \qquad \mbox{and} \qquad
  \condsigma(\tau)
  \;=\;
  \frac{1}{\sqrt{\frac{1}{\priorsigma^2}+\sum_{i=1}^k\frac{1}{\sigma_i^2+\tau^2}}}\mbox{,}
\end{equation}
and the logarithmic marginal likelihood turns out as
\begin{eqnarray}
  \log\bigl(p(\vec{y}|\tau,\vec{\sigma})\bigr) 
  &=&
  -\frac{1}{2}\Biggl(k\log(2\pi) +
    \log\bigl(\priorsigma^2\bigr)+\sum_{i=1}^k\log\bigl(\sigma_i^2\!+\!\tau^2\bigr)
  \nonumber \\
  && \qquad 
+ \frac{\bigl(\priormu - \condmu(\tau)\bigr)^2}{\priorsigma^2} + \sum_{i=1}^k \frac{\bigl(y_i - \condmu(\tau)\bigr)^2}{\sigma_i^2+\tau^2}
+ \log\biggl(\frac{1}{\priorsigma^2}+\sum_{i=1}^k\frac{1}{\sigma_i^2\!+\!\tau^2}\biggr) \Biggr)\mbox{.}
  \label{eqn:marginal2b}
\end{eqnarray}
  Note that, comparing equations~(\ref{eqn:marginal1c}) and
  (\ref{eqn:marginal2b}) (as well as (\ref{eqn:conditionalmoments1})
  and (\ref{eqn:conditionalmoments2})), as expected, use of the
  uniform prior constitutes the limiting case of large prior
  uncertainty ($\priorsigma\rightarrow\infty$).

\subsection{Conditional effect posteriors}\label{sec:conditional}
  As long as a uniform or normal prior for the effect~$\mu$ is used,
  the effect's conditional posterior distribution for a given
  heterogeneity value, $p(\mu|\tau,\vec{y},\vec{\sigma})$, again is
  normal with mean~$\condmu(\tau)$ and standard deviation
  $\condsigma(\tau)$ as given in equations
  (\ref{eqn:conditionalmoments1}) or (\ref{eqn:conditionalmoments2}),
  respectively \citep{BDA3rd}.

  Note that the conditional posterior moments
  (\ref{eqn:conditionalmoments1}) are also commonly utilized in
  frequentist fixed-effect and random-effects meta-analyses.  The
  mean~$\condmu(\tau)$ constitutes the \emph{conditional maximum
    likelihood estimate} (of~$\mu$), conditional on a particular
  amount of heterogeneity~$\tau$, while $\condsigma(\tau)$ gives the
  corresponding (conditional) standard error.  Plugging in
  $\tau\!=\!0$ yields the \emph{fixed-effect estimate} of~$\mu$,
  while a value $\tau>0$ yields a \emph{random-effects estimate}
  \citep[Sec.~6]{HedgesOlkin}; see also
  Section~\ref{sec:frequentistConnection} below for an example.

\subsection{Marginal and joint posterior}\label{sec:MargJoint}
  Having derived the marginal likelihood
  $p(\vec{y}|\tau,\vec{\sigma})$ in Section~\ref{sec:marginal}, the
  (one-dimensional) mar\-gin\-al posterior density of $\tau$ may be
  computed (up to a normalizing constant) by multiplication with the
  heterogeneity prior
  \begin{equation}\label{eqn:heteroMarginal}
    p(\tau|\vec{y},\vec{\sigma}) \;\propto\; p(\vec{y}|\tau,\vec{\sigma}) \times p(\tau)\mbox{.}
  \end{equation}
  This feature was one of the reasons for specifying the priors for
  $\mu$ and $\tau$ as independent (see
  Section~\ref{sec:priorGeneral}).  One-dimensional integration can
  now easily be done numerically for arbitrary priors~$p(\tau)$, as
  long as the resulting posterior is proper.

  The effect's conditional posterior
  $p(\mu|\tau,\vec{y},\vec{\sigma})$ (see
  Section~\ref{sec:conditional}) is of particular interest, since the
  joint posterior may be re-ex\-press\-ed in terms of the conditional
  as
  \begin{equation}\label{eqn:mixture}
    p(\mu,\tau|\vec{y},\vec{\sigma}) 
    \;=\; p(\mu|\tau,\vec{y},\vec{\sigma}) \times p(\tau|\vec{y},\vec{\sigma})\mbox{.}
  \end{equation}
  In this formulation, it becomes obvious that the effect's marginal
  distribution is a continuous \emph{mixture distribution}, in which
  the normal conditionals $p(\mu|\tau,\vec{y},\vec{\sigma})$ are mixed
  via the marginal $p(\tau|\vec{y},\vec{\sigma})$ with
  \begin{equation}
    p(\mu|\vec{y},\vec{\sigma}) 
    \;=\; \int p(\mu,\tau|\vec{y},\vec{\sigma}) \,\mathrm{d}\tau
    \;=\; \int p(\mu|\tau, \vec{y},\vec{\sigma}) \times p(\tau|\vec{y},\vec{\sigma}) \,\mathrm{d}\tau
  \end{equation}
  \citep{Seidel2010,Lindsay}.  This expression allows for easy
  numerical approximation of posterior integrals of interest.  For
  example, the marginal distribution of the effect~$\mu$ (the normal
  mixture) may be approximated by using a discrete grid of
  $\tau$~values and summing up the normal conditionals using weights
  defined through $\tau$'s marginal density:
  \begin{equation}\label{eqn:gridApprox}
    p(\mu) 
    \;=\; \int p(\mu | \tau)\, p(\tau)\, \differential \tau 
    \;\approx\; \sum_j p(\mu | \tau_j) \, w_j \mbox{,}
  \end{equation}
  where the set of $\tau_j$ is appropriately chosen and corresponding
  ``weights'' $w_j$ (with $\sum_j w_j = 1$) are based on the marginal
  $p(\tau)$.  With that, it is now relatively straightforward to work
  with the joint distribution, derive marginals, moments, implement
  Monte Carlo integration, and so on.  A general prescription of how
  to approach a discrete approximation as sketched in
  (\ref{eqn:gridApprox}) while keeping the accuracy under control is
  given by the \textsc{direct} algorithm described by
  \citet{RoeverFriede2017}. A few more technical details are also
  given in Section~\ref{sec:bayesmetaDetails} and
  Appendix~\ref{sec:directDetails} below.

\subsection{Predictive distribution}\label{sec:predictive}
  The predictive distribution expresses the posterior knowledge about
  a ``future'' observation, i.e., an additional draw~$\theta_{k+1}$
  from the underlying population of studies.  This is commonly of
  interest in order to judge the amount of heterogeneity relative to
  the estimation uncertainty
  \citep{RileyHigginsDeeks2011,GuddatEtAl2012,BenderEtAl2014}, or for
  extrapolation in the design and analysis of future studies
  \citep{SchmidliEtAl2014}.  Technically, the predictive distribution
  $p(\theta_{k+1}|\vec{y},\vec{\sigma})$ is similar to the marginal
  distribution of the effect~$\mu$ (see previous section).
  Conditionally on a given heterogeneity~$\tau$, and for the uniform
  or normal effect prior, the predictive distribution again is normal
  with moments
  \begin{equation}
    \expect[\theta_{k+1} \,|\, \tau, \vec{y}, \vec{\sigma}]
    \;=\; \condmu(\tau)
    \qquad \mbox{and} \qquad
    \var(\theta_{k+1} \,|\, \tau, \vec{y}, \vec{\sigma})
    \;=\; \condsigma^2(\tau) + \tau^2.
  \end{equation}

\subsection{Shrinkage estimates of study-specific means}\label{sec:shrinkage}
  Sometimes it is of interest to also infer the posterior
  distributions of the study-specific parameters~$\theta_j$. These may
  e.g.\ be in the focus if a meta-analysis is performed in order
  support the analysis of a particular study by borrowing strength
  from a number of related studies
  \citep{BDA3rd,SchmidliEtAl2014,WandelNeuenschwanderRoeverFriede2017}.
  Conditionally on a particular heterogeneity value~$\tau$, these
  distributions are again normal with moments given by
  \begin{eqnarray}\label{eqn:shrinkExpect}
    \expect[\theta_j \,|\, \tau, \vec{y}, \vec{\sigma}]
    &=& \frac{\frac{1}{\sigma_j^2}y_j+\frac{1}{\tau^2}\condmu(\tau)}{\frac{1}{\sigma_j^2}+\frac{1}{\tau^2}}
    \\ \label{eqn:shrinkVar}
    \var(\theta_j \,|\, \tau, \vec{y}, \vec{\sigma})
    &=& \frac{1}{\frac{1}{\sigma_j^2}+\frac{1}{\tau^2}} + \Biggl(\frac{\frac{1}{\tau^2}}{\frac{1}{\sigma_j^2}+\frac{1}{\tau^2}} \condsigma\Biggr)^2 
  \end{eqnarray}
  \citep[Sec.~5.5]{BDA3rd}.  These expressions illustrate the
  \emph{shrinkage} of posterior estimates towards the common mean as
  a function of the heterogeneity.  Analogously to the effect's
  posterior and predictive distribution, these conditional moments
  again allow to approximate each individual~$\theta_i$'s marginal
  posterior distribution via a discrete mixture to marginalize over
  the heterogeneity.

\subsection{Credible intervals}\label{sec:CIs}
  Credible intervals derived from a posterior probability distribution
  may be computed e.g.\ using the distribution's $\frac{\alpha}{2}$
  and $(1\!-\!\frac{\alpha}{2})$ quantiles. However, such a simple
  \emph{central} interval may not necessarily be the most sensible
  summary of a posterior distribution, especially if it is skewed or
  extends to the boundary of its parameter space. In such cases, it
  usually makes more sense to consider the \emph{highest posterior
    density (HPD) region}, i.e., a $(1\!-\!\alpha)$ credible region
  enclosing the $(1\!-\!\alpha)$ posterior probability where the
  posterior density is largest \citep{BDA3rd}. Such a region may be
  disjoint and hard to determine, but closely related (and identical
  for unimodal distributions) is the \emph{shortest credible
    interval}. Both types of intervals, central and shortest, will be
  considered in the following.

\subsection[Posterior predictive checks and p-values]{Posterior predictive checks and $p$-values}\label{sec:PostPred}
  Posterior predictive model checks allow to investigate the fit of a
  model to a given data set
  \citep{GelmanMengStern1996,Gelman2003,BDA3rd}. The consistency of
  data and model is explored by comparing the actual data to data sets
  \emph{predicted} via the posterior distribution. The comparison is
  usually done graphically, or via suitable summary statistics of
  actual and predicted data; a discrepancy then is an indicator of a
  poor model fit.

  If the summary statistic is one-dimensional, then the comparison may
  be formalized by focusing on the fractions of predicted values above
  or below the actually observed value. This leads to the concept of
  \emph{posterior predictive $p$\mbox{-}values}, which are closely related
  to classical $p$\mbox{-}values
  \citep{Meng1994,BerkhofEtAl2000,Gelman2013,Wasserstein2016}.
  Posterior predictive $p$\mbox{-}values have been applied and advocated in a
  range of contexts, including e.g.\ educational testing
  \citep{SinharayJohnsonStern2006}, metrology
  \citep{KackerForbesKesselSommer2008}, psychology
  \citep{VanDeSchootEtAl2014} and biology
  \citep{ChambertRotellaHiggs2014}.

  In the context of the NNHM, posterior predictive checks are useful,
  as they allow to investigate certain hypotheses of interest, like
  for example $\mu \geq 0$, $\tau=0$ or $\theta_i = 0$.  The posterior
  predictive distribution \emph{conditional on a particular
    hypothesis} may then be explored in order to derive a
  corresponding posterior predictive $p$\mbox{-}value.  The choice of a
  suitable summary statistic however may still pose a challenge.  The
  posterior predictive checks here are implemented via Monte Carlo
  sampling, therefore parts of these procedures are computationally
  expensive.

\subsection[How the bayesmeta() function works internally]{How the \code{bayesmeta()} function works internally}\label{sec:bayesmetaDetails}
  The \code{bayesmeta()} function utilizes the fact that in the
  context of the NNHM the resulting posterior is only 2-dimensional
  (for now ignoring the $\theta_i$ parameters) and may be expressed as
  a mixture distribution (see~(\ref{eqn:mixture})) where the
  heterogeneity's marginal $p(\tau|\vec{y},\vec{\sigma})$ is known,
  and the effect's conditionals $p(\mu|\tau,\vec{y},\vec{\sigma})$ are
  all of a normal form. This setup allows to approximate the effect
  marginal by a discrete mixture (see~(\ref{eqn:gridApprox})) while
  keeping the accuracy under control; the accuracy requirements are
  formulated via the \textsc{direct} algorithm's two tuning parameters
  ($\delta$~and~$\epsilon$) \citep{RoeverFriede2017}.

  An example of joint and marginal posterior densities of the two
  parameters is illustrated in Figure~\ref{fig:plot} below (see
  page~\pageref{fig:plot}).  The joint posterior density (top right)
  is easily evaluated based on likelihood and prior density, both of
  which are available in analytical form (see Sections~\ref{sec:prior}
  and~\ref{sec:likelihood}).  The heterogeneity's marginal density
  (bottom right) is also easily computed, based on marginal likelihood
  and prior (see (\ref{eqn:heteroMarginal})); only its normalizing
  constant needs to be computed numerically (using the
  \code{integrate()} function available in \proglang{R}). The CDF is
  also computed using numerical integration, and the quantile function
  is evaluated using again the CDF and inverting it via \proglang{R}'s
  \code{uniroot()} root-finding function.

  Now the effect's marginal density (bottom left panel of
  Figure~\ref{fig:plot}) is approximated by a mixture of a finite
  number of normal distributions. In terms of equation
  (\ref{eqn:gridApprox}), what is required is a finite set of support
  points~$\tau_j$, the parameters (means and standard deviations) of
  the associated normal conditionals $p(\mu|\tau_j)$, and the
  corresponding weights~$w_j$. These are all determined using the
  \textsc{direct} algorithm, and in the \code{bayesmeta()} output (see
  the following section) one can find these in the
  ``\code{...\$support}'' element.  In the example shown in
  Figure~\ref{fig:plot}, the effect marginal is based on a
  17-component normal mixture; this number of components is sufficient
  to bound the discrepancy between actual marginal and mixture
  approximation to amount to a Kullback-Leibler divergence
  below~\mbox{$\delta\!=\!1\%$}. The desired accuracy can be
  pre-specified via the ``\code{delta}'' and ``\code{epsilon}''
  arguments \citep{RoeverFriede2017}.

  Computations related to such discrete, finite mixtures are
  relatively straightforward; density and CDF are linear combination
  of the components' (normal) densities and CDFs, random number
  generation is simple, and moments are also easily derived
  \citep{Seidel2010,Lindsay}. A few more details on the implementation
  are given in Appendix~\ref{sec:directDetails}. Many of the internal
  computations heavily rely on numerical integration, root-finding and
  optimization via \proglang{R}'s \code{integrate()},
  \code{uniroot()}, \code{optimize()} and \code{optim()} functions.
  Accuracy of the eventual implementation is confirmed 
  using simulations in Appendix~\ref{sec:calibration}.

\newpage    
\section[Using the bayesmeta package]{Using the \pkg{bayesmeta} package}\label{sec:usage}
\subsection{General}
  Before proceeding to an exemplary analysis, we will first introduce
  an example data set and go through the common procedure of effect
  size derivation step-by-step. This will serve to introduce some
  context and generate a set of estimates~($y_i$) and associated
  standard errors~($\sigma_i$); the subsequent section will then pick
  up the analysis from that starting point.

\subsection{Example data: a systematic review in immunosuppression}\label{sec:CrinsData}
  Interleukin-2 receptor antagonists (IL-2RA) are commonly used as
  part of immunosuppressive therapy after organ transplantation.
  Treatment strategies and responses are different for adults and
  children, and it was of interest to investigate the effectiveness of
  IL-2RA in preventing acute rejection (AR) events following liver
  transplantation in paediatric patients. A systematic literature
  review was performed, and six controlled studies were found
  reporting on the occurrence of AR events in paediatric liver
  transplant recipients \citep{CrinsEtAl2014}.

  \begin{table}[b]
  \begin{center}
    \caption{\label{tab:2x2}The general setup of a $2\!\times\!2$
      contingency table for dichotomous outcomes (left) and a concrete example from the
      paediatric liver transplantation data set (right). Note that one
      of the three data columns is redundant here, as it may be derived
      from the remaining two.} \vspace{1ex}
    \begin{tabular}{lrrr}
      \toprule
                        & \multicolumn{2}{c}{event} & \\ \addlinespace[-0.3ex]
      \cmidrule(lr){2-3}
                        & yes & no & total \\
      \midrule
      treatment  & $a$ & $b$ & $n_1\!=\!a\!+\!b$ \\
      control    & $c$ & $d$ & $n_2\!=\!c\!+\!d$ \\
      \bottomrule
    \end{tabular}
    \hspace{8ex}
    \begin{tabular}{lrrr}
      \toprule
                        & \multicolumn{2}{c}{AR event} & \\ \addlinespace[-0.3ex]
      \cmidrule(lr){2-3}
                        & yes & no & total \\
      \midrule
      IL-2RA patients   & $14$ & $47$ & $61$ \\
      control patients  & $15$ &  $5$ & $20$ \\
      \bottomrule
    \end{tabular}
  \end{center}
  \end{table}

  The binary data on AR events from each of the six studies may be
  summarized in a $2\!\times\!2$-table as shown in
  Table~\ref{tab:2x2}. The data shown here come from the earliest of
  the studies found in the review \citep{HeffronEtAl2003}.  Here one
  can already see that the treatment appears to be effective, as
  roughly only a quarter of patients in the IL-2RA group experienced
  an AR event, compared to three quarters in the control group.

  In order to compare the effect magnitude between different studies,
  a common effect measure is computed from each contingency table (for
  each study~$i$). One such measure is the logarithmic odds ratio
  (log-OR), comparing the odds of an event in treatment- and
  control-groups. The log-OR estimate is given by
  $y_i=\log\Bigl(\frac{a/b}{c/d}\Bigr)$, where $a$ to $d$ are the
  event counts as defined in Table~\ref{tab:2x2}; the corresponding
  standard error is
  $\sigma_i=\sqrt{\frac{1}{a}+\frac{1}{b}+\frac{1}{c}+\frac{1}{d}}$.
  In the above example, the odds ratio is
  $\frac{14/47}{15/5}=\frac{14}{141}\approx 0.10$; we have
  $y_1=\log\Bigl(\frac{14/47}{15/5}\Bigr) = -2.31$ and
  $\sigma_1=\sqrt{\frac{1}{14}+\frac{1}{47}+\frac{1}{15}+\frac{1}{5}}
  = 0.60$.  A wide range of other measures is available for
  contingency tables as well as other types of study outcomes; for
  example, in the present case one might alternatively be interested
  in (logarithmic) \emph{relative risks (RR)}
  ($\log\Bigl(\frac{a/(a+b)}{c/(c+d)}\Bigr)$) instead of the log-ORs
  \citep{HedgesOlkin,HartungKnappSinha,BorensteinEtAl,Viechtbauer2010,HigginsGreen,Deeks2002}.
  The original data and derived log-ORs for all six studies from the
  systematic review are shown in Table~\ref{tab:CrinsData}.

  \begin{table}
  \begin{center}
    \caption{\label{tab:CrinsData}Data from the immunosuppression
      example. Each row here summarizes a $2\!\times\!2$ contingency
      table, the last two columns show the correponding derived
      log-ORs~($y_i$) and their associated standard
      errors~($\sigma_i$).} \vspace{1ex}
    \defcitealias{HeffronEtAl2003}{Heffron \textit{et~al.}}
    \defcitealias{GibelliEtAl2004}{Gibelli \textit{et~al.}}
    \defcitealias{SchullerEtAl2005}{Schuller \textit{et~al.}}
    \defcitealias{GanschowEtAl2005}{Ganschow \textit{et~al.}}
    \defcitealias{SpadaEtAl2006}{Spada \textit{et~al.}}
    \defcitealias{GrasEtAl2008}{Gras \textit{et~al.}}
    \small
    \begin{tabular}{cllrrrrrr} 
      \toprule
      \multicolumn{3}{c}{study} & \multicolumn{2}{c}{IL-2RA group} & \multicolumn{2}{c}{control group} & \multicolumn{2}{c}{log-OR}\\ 
      \cmidrule(lr){1-3} \cmidrule(lr){4-5} \cmidrule(lr){6-7} \cmidrule(lr){8-9}
      $i$ & author & year & events ($a_i$) & total ($n_{1;i}$) & events ($c_i$) & total ($n_{2;i}$) & $y_i$ & $\sigma_i$ \\
      \midrule
      1 & \citetalias{HeffronEtAl2003}  & 2003 & 14 & 61 & 15 & 20 & $-2.31$ & $0.60$ \\
      2 & \citetalias{GibelliEtAl2004}  & 2004 & 16 & 28 & 19 & 28 & $-0.46$ & $0.56$ \\
      3 & \citetalias{SchullerEtAl2005} & 2005 &  3 & 18 &  8 & 12 & $-2.30$ & $0.88$ \\
      4 & \citetalias{GanschowEtAl2005} & 2005 &  9 & 54 & 29 & 54 & $-1.76$ & $0.46$ \\
      5 & \citetalias{SpadaEtAl2006}    & 2006 &  4 & 36 & 11 & 36 & $-1.26$ & $0.64$ \\
      6 & \citetalias{GrasEtAl2008}     & 2008 &  0 & 50 &  3 & 34 & $-2.42$ & $1.53$ \\
      \bottomrule
    \end{tabular}
  \end{center}
  \end{table}

  The transplantation data set is also contained in the
  \pkg{bayesmeta} package; the data need to be loaded via the
  \code{data()} function:
\begin{CodeChunk}
\begin{CodeInput}
R> require("bayesmeta")
R> data("CrinsEtAl2014")
R> CrinsEtAl2014
\end{CodeInput}
\end{CodeChunk}
  Effect sizes and standard errors can be calculated from the plain
  count data either by implementing the corresponding formulas (see
  above), or, much easier and recommended, by using e.g.\ the
  \pkg{metafor} package's \code{escalc()} function:
\begin{CodeChunk}
\begin{CodeInput}
R> require("metafor")
R> crins.es <- escalc(measure="OR",
+    ai=exp.AR.events, n1i=exp.total, ci=cont.AR.events, n2i=cont.total,
+    slab=publication, data=CrinsEtAl2014)
R> crins.es
\end{CodeInput}
\end{CodeChunk}
  One can see that the \code{escalc()} function uses a terminology
  analogous to that in Table~\ref{tab:2x2} to interface with binary
  outcome data; the ``\code{ai}''~input argument corresponds to the
  $a_i$ table entries (number~$a$ of events in the treatment group for
  each study~$i$), and so on. The output of the \code{escalc()}
  function (here: the data frame named ``\code{crins.es}'') will then
  be the original data along with two additional columns named
  ``\code{yi}'' and ``\code{vi}'' containing the calculated effect
  sizes ($y_i$) and the \emph{squared} (!)  standard errors
  ($\sigma_i^2$), respectively.

  Note that for computing the $6$th study's log-OR (see
  Table~\ref{tab:CrinsData}), a \emph{continuity correction} was
  necessary, because one of the contingency table entries was zero
  \citep{SweetingSuttonLambert2004}. For more details on effect size
  calculation and default behaviour, see also \citet{Viechtbauer2010}
  or the \code{escalc()} function's online documentation.

\subsection{Performing a Bayesian random-effects meta-analysis}
\subsubsection[The bayesmeta() function]{The \code{bayesmeta()} function}
  In order to perform a random-effects meta-analysis, we need to
  specify the data, as well as the prior for the unknown parameters
  $\mu$ and $\tau$ (see Section~\ref{sec:prior}).  For the
  effect~$\mu$ we are restricted to normal or uniform priors; here we
  use a vague prior centered at $\priormu\!=\!0$, which corresponds to
  an OR of~1, i.e., no effect. The prior standard deviation we set
  to $\priorsigma\!=\!4$, corresponding to the vague \emph{unit
    information prior} (see Section~\ref{sec:effectPrior}).  For the
  heterogeneity, we use a half-normal prior with scale~$0.5$,
  confining the a~priori expected heterogeneity to $\tau \leq 0.98$
  with $95\%$ probability (i.e., allowing for ``fairly extreme''
  values with only about $5\%$ prior probability).

  With the log-ORs computed as in the previous section, we can now
  execute the analysis using the following call
\begin{CodeChunk}
\begin{CodeInput}
R> ma01 <- bayesmeta(y = crins.es[,"yi"], sigma = sqrt(crins.es[,"vi"]),
+    labels = crins.es[,"publication"], mu.prior.mean = 0, mu.prior.sd = 4,
+    tau.prior = function(t){dhalfnormal(t,scale=0.5)})
\end{CodeInput}
\end{CodeChunk}
  The first three arguments pass the data (vectors of estimates~$y_i$
  and standard errors~$\sigma_i$) and (optionally) a vector of
  corresponding study labels to the \code{bayesmeta()} function.  Note
  that the \pkg{metafor} package's \code{escalc()} function returned
  variances (i.e., \emph{squared} standard errors), while the
  \code{bayesmeta()} function's ``\code{sigma}'' argument requires the
  standard errors (i.e., the square root of the variances); hence the
  additional square-root-transformation here.  The following arguments
  specify the prior mean and standard deviation of the (normal) prior
  for the effect~$\mu$. Finally, the last argument specifies the prior
  for the heterogeneity~$\tau$. While for the effect prior we are
  restricted to using normal or improper uniform priors, the
  heterogeneity prior can be of essentially any type. Specification of
  the heterogeneity prior works via specification of its \emph{prior
    density function}. While this type of argument specification is
  somewhat unusual, it is reasonably straightforward, as one can see
  above. The \code{dhalfnormal()} function here is the half-normal
  distribution's density function; see also the corresponding online
  help (e.g.\ via entering ``\code{?dhalfnormal}'' in \proglang{R}).

  Retrieving and processing the ``\code{yi}'' and ``\code{vi}''
  elements (as well as study labels, if available) from an
  \code{escalc()} result in general is not complicated, and the
  \code{bayesmeta()} function can also do this automatically for any
  \code{escalc()} output, including the many types of effect sizes
  that are available \citep{Viechtbauer2010}.  Using simply the
  \code{escalc()} function's output as an input, the identical result
  can be achieved by calling
\begin{CodeChunk}
\begin{CodeInput}
R> ma01 <- bayesmeta(crins.es, mu.prior.mean = 0, mu.prior.sd = 4,
+    tau.prior = function(t){dhalfnormal(t,scale=0.5)})
\end{CodeInput}
\end{CodeChunk}
  The \code{bayesmeta()} computations may take up to a few seconds,
  but with that the main calculations are done, and the essential
  results are stored in the generated object of class
  ``\code{bayesmeta}'' (here named ``\code{ma01}''). One can inspect
  the results by printing the returned object:
\begin{CodeChunk}
\begin{CodeInput}
R> ma01
\end{CodeInput}
\begin{CodeOutput}
 'bayesmeta' object.

6 estimates:
Heffron (2003), Gibelli (2004), Schuller (2005), Ganschow (2005), 
Spada (2006), Gras (2008)

tau prior (proper):
function(t){dhalfnormal(t,scale=0.5)}

mu prior (proper):
normal(mean=0, sd=4)

ML and MAP estimates:
                    tau        mu
ML joint     0.32581341 -1.578262
ML marginal  0.46441292 -1.578003
MAP joint    0.08690907 -1.559376
MAP marginal 0.24531385 -1.569122

marginal posterior summary:
                tau         mu
mode      0.2453139 -1.5691216
median    0.3445022 -1.5734823
mean      0.3810562 -1.5764366
sd        0.2593672  0.3295298
95% lower 0.0000000 -2.2312306
95% upper 0.8607305 -0.9264079

(quoted intervals are shortest credible intervals.)
\end{CodeOutput}
\end{CodeChunk}
  One can see that the analysis was based on $k\!=\!6$ studies, that
  both parameters' priors were found to be proper, and
  maximum-likelihood (ML) as well as maximum-a-posteriori (MAP) values
  are quoted. Probably most interestingly, under ``marginal posterior
  summary'' one can find summary statistics describing the marginal
  posterior distributions of heterogeneity~($\tau$) and
  effect~($\mu$), which may often be the most relevant figures.  The
  resulting posterior median and 95\% credible interval for the
  effect~$\mu$ here are at a log-OR of $-1.57$ [$-2.23$, $-0.93$];
  this information may eventually constitute the essential result
  in many cases.

\subsubsection[The forestplot() function]{The \code{forestplot()} function}
\begin{figure}[t]
    \begin{center}
      \includegraphics[width=0.9\linewidth]{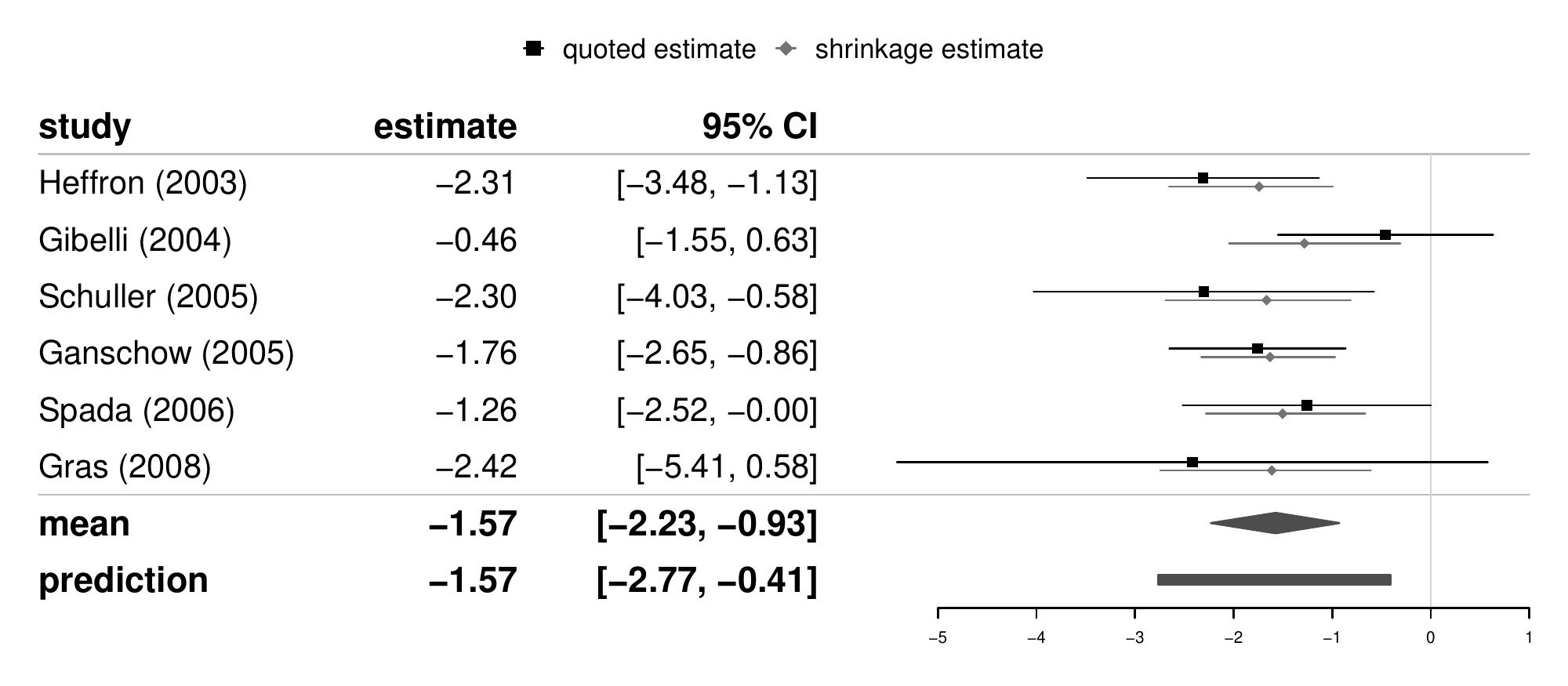}
      \caption{\label{fig:forestplot01} A forest plot, generated using
        the \code{forestplot()} function with default settings,
        showing the input data, effect estimate, prediction interval
        and shrinkage estimates.}
    \end{center}
  \end{figure}

  To illustrate data and results, one can use the \code{forestplot()}
  function. This function is actually a \code{bayesmeta}-specific
  method based on the \pkg{forestplot} package's generic
  \code{forestplot()} function \citep{R:forestplot}.  In its simplest
  form, it may be used as
\begin{CodeChunk}
\begin{CodeInput}
R> forestplot(ma01)
\end{CodeInput}
\end{CodeChunk}
  Figure~\ref{fig:forestplot01} shows the \code{forestplot()}
  function's default output for the example analysis.
  In the figure one can see all estimates~$y_i$ along with
  95\%~intervals based on the provided standard errors~$\sigma_i$.  At
  the bottom, 95\% credible intervals for the effect and for the
  predictive distribution are shown
  \citep{LewisClarke2001,GuddatEtAl2012}. Next to each of the quoted
  estimates (as specified through $y_i$ and~$\sigma_i$), the
  \emph{shrinkage intervals} for the study-specific effects~$\theta_i$
  are also shown in grey; these illustrate the posterior of each
  individual study's true effect (see equation~(\ref{eqn:NNHM2}) and
  Sec.~\ref{sec:shrinkage}). The forest plot can be customized in many
  ways; one can add columns to the table, change axis scaling and
  labels, omit shrinkage or prediction intervals, etc. For all the
  options see the online documentation for the
  \code{forestplot.bayesmeta()} method.

\subsubsection[The plot() function]{The \code{plot()} function}
\begin{figure}[h!]
    \begin{center}
      \includegraphics[width=0.9\linewidth]{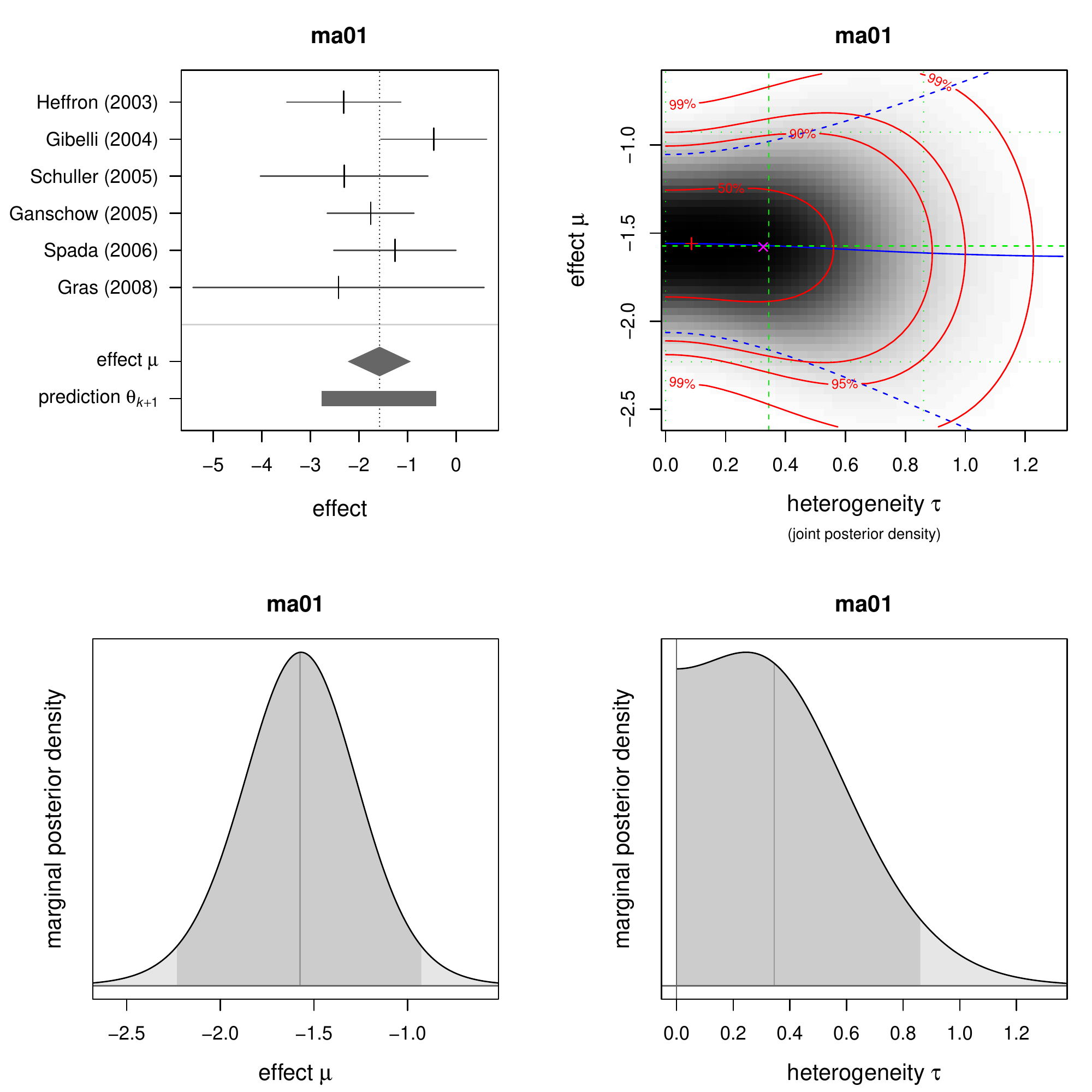}
      \caption{\label{fig:plot}The four plots generated via the
        \code{plot()} function. The top left plot is a simple forest
        plot showing estimates and 95\% intervals illustrating the
        input data ($y_i$ and $\sigma_i$) along with the estimated
        mean effect~$\mu$ and a prediction interval for the
        effect~$\theta_{k+1}$ in a future study. The top right plot
        illustrates the joint posterior density of
        heterogeneity~$\tau$ and effect~$\mu$, with darker shading
        corresponding to higher probability density. The red lines
        indicate (approximate) 2\mbox{-}dimensional credible regions,
        and the green lines show marginal posterior medians and 95\%
        credible intervals. The blue lines show the conditional
        posterior mean effect~$\condmu(\tau)$ as a function of the
        heterogeneity~$\tau$ along with a 95\% interval based on its
        conditional standard error~$\condsigma(\tau)$ (see also
        Section~\ref{sec:marginal}). The red cross~($+$) indicates the
        posterior mode, while the pink cross~($\times$) shows the ML
        estimate. The two bottom plots show the marginal posterior
        densities of effect~$\mu$ and heterogeneity~$\tau$. 95\%
        credible intervals are indicated with a darker shading, and
        the posterior median is shown by a vertical line.}
    \end{center}\vspace{4ex} 
  \end{figure}

  The analysis output may be inspected more closely using the
  \code{plot()} function:
\begin{CodeChunk}
\begin{CodeInput}
R> plot(ma01)
\end{CodeInput}
\end{CodeChunk}
  The output for our example is shown in Figure~\ref{fig:plot};
  in particular, the joint and marginal posterior distributions are
  illustrated in detail. Prior densities may be superimposed by using
  the ``\code{prior=TRUE}'' argument, and axis ranges may also be
  specified manually; see also the online help for the
  \code{plot.bayesmeta()} method.

\subsubsection[Elements of the bayesmeta() output]{Elements of the \code{bayesmeta()} output}
  It is possible to access the joint and marginal densities shown in
  Figure~\ref{fig:plot} (and more) directly from the
  \code{bayesmeta()} output.  As usual for an object returned from a
  non-trivial analysis function, the result of a \code{bayesmeta()}
  call is a \code{list} object of class ``\code{bayesmeta}''
  containing a number of further individual objects. One can check the
  complete listing of available entries in the online
  documentation. For example, there is the ``\code{...\$summary}''
  entry giving some basic summary statistics:
\begin{CodeChunk}
\begin{CodeInput}
R> ma01$summary
\end{CodeInput}
\begin{CodeOutput}
                tau         mu      theta
mode      0.2453139 -1.5691214 -1.5632732
median    0.3445023 -1.5734819 -1.5701653
mean      0.3810562 -1.5764365 -1.5764365
sd        0.2593672  0.3295301  0.5671855
95% lower 0.0000000 -2.2311251 -2.7661319
95% upper 0.8607193 -0.9263075 -0.4072970
\end{CodeOutput}
\end{CodeChunk}
  Some of these we already saw in the output when simply printing the
  object (see above). The additional third column here shows summary
  statistics for the predictive distribution of a `future'
  study~($\theta_{k+1}$). One can also access the original data (the
  $y_i$ and $\sigma_i$) in the ``\code{...\$y}'' and
  ``\code{...\$sigma}'' entries, or the study labels and the total
  number of studies~($k$) in the ``\code{...\$labels}'' and
  ``\code{...\$k}'' entries.

  Most importantly, some of the elements are \code{function}s allowing
  to access and evaluate the various posterior distributions. For
  example, the posterior density can be accessed via the
  ``\code{...\$dposterior()}'' function; this function has a
  ``\code{mu}'' or a ``\code{tau}'' argument, specifying either of
  these results in a marginal density, and specifying both gives the
  joint density.  So a simple plot of the effect's marginal posterior
  density can be generated by
\begin{CodeChunk}
\begin{CodeInput}
R> x <- seq(-3, 0.5, length=200)
R> plot(x, ma01$dposterior(mu=x), type="l",
+    xlab="effect", ylab="posterior density")
R> abline(h=0, v=0, col="grey")
\end{CodeInput}
\end{CodeChunk}
  In order to calculate the posterior probability of a
  non-beneficial effect 
  (\mbox{$\prob(\mu>0|\vec{y},\vec{\sigma})$} $=$ \mbox{$1-\prob(\mu\leq 0|\vec{y},\vec{\sigma})$}), 
  one needs to evaluate the
  marginal posterior cumulative distribution function (CDF). This is
  provided via the ``\code{...\$pposterior()}'' function:
\begin{CodeChunk}
\begin{CodeInput}
R> 1 - ma01$pposterior(mu=0)
\end{CodeInput}
\begin{CodeOutput}
[1] 6.187343e-05
\end{CodeOutput}
\end{CodeChunk}
  Or one can also plot the complete CDF using the following code:
\begin{CodeChunk}
\begin{CodeInput}
R> x <- seq(-3, 0.5, length=200)
R> plot(x, ma01$pposterior(mu=x), type="l",
+    xlab="effect", ylab="posterior CDF")
R> abline(h=0:1, v=0, col="grey")
\end{CodeInput}
\end{CodeChunk}
  The same works also for the heterogeneity parameter~$\tau$; in order
  to derive for example the posterior probability for a ``fairly
  extreme'' heterogeneity ($\tau > 1$), one simply needs to supply the
  ``\code{tau}'' parameter instead:
\begin{CodeChunk}
\begin{CodeInput}
R> 1 - ma01$pposterior(tau=1)
\end{CodeInput}
\begin{CodeOutput}
[1] 0.02097488
\end{CodeOutput}
\end{CodeChunk}
  so the posterior probability is at $2.1\%$ here. The quantile
  function (inverse CDF) is also available in the
  ``\code{...\$qposterior()}'' function; in order to derive for
  example a 99\% upper limit on the heterogeneity parameter, one needs
  to evaluate
\begin{CodeChunk}
\begin{CodeInput}
R> ma01$qposterior(tau.p=0.99)
\end{CodeInput}
\begin{CodeOutput}
[1] 1.109186
\end{CodeOutput}
\end{CodeChunk}
  so the 99\% upper limit would here be at $\tau=1.11$.

  In many cases it is useful to use Monte Carlo simulation to derive
  other non-trival quantities from the posterior distribution. One can
  generate samples from the posterior distribution using the
  ``\code{...\$rposterior()}'' function. A call of
\begin{CodeChunk}
\begin{CodeInput}
R> ma01$rposterior(n=5)
\end{CodeInput}
\begin{CodeOutput}
            tau        mu
[1,] 0.23423926 -1.380271
[2,] 0.28630556 -1.442691
[3,] 0.04402682 -1.610052
[4,] 0.83672662 -1.550758
[5,] 0.18981184 -1.803012
\end{CodeOutput}
\end{CodeChunk}
  will generate a sample of 5~draws from the joint (bivariate)
  posterior distribution of~$\tau$ and~$\mu$. If one is only
  interested in the marginal distribution of~$\mu$, it is
  (substantially!) more efficient to omit the $\tau$ draws and use
\begin{CodeChunk}
\begin{CodeInput}
R> ma01$rposterior(n=5, tau.sample=FALSE)
\end{CodeInput}
\begin{CodeOutput}
[1] -2.184596 -1.876711 -1.514224 -1.384694 -1.567397
\end{CodeOutput}
\end{CodeChunk}
  to generate a vector of $\mu$ values only.

  For example, suppose that we assume a rate of AR events of
  $p_c=50\%$ for the control group, and we are interested in the
  implied \emph{risk difference} based on our analysis. The risk
  difference is simply $p_t-p_c$, where $p_t$ is the event rate in the
  treatment (IL-2RA) group. To determine the distribution of the risk
  difference we can now simply use Monte Carlo sampling and run
\begin{CodeChunk}
\begin{CodeInput}
R> prob.control <- 0.5
R> logodds.control <- log(prob.control / (1 - prob.control))
R> logodds.treat <- (logodds.control 
+                    + ma01$rposterior(n=10000, tau.sample=FALSE))
R> prob.treat <- exp(logodds.treat) / (1 + exp(logodds.treat))
R> riskdiff <- (prob.treat - prob.control)
R> median(riskdiff)
\end{CodeInput}
\begin{CodeOutput}
[1] -0.3284975
\end{CodeOutput}
\begin{CodeInput}
R> quantile(riskdiff, c(0.025, 0.975))
\end{CodeInput}
\begin{CodeOutput}
      2.5%      97.5% 
-0.4028368 -0.2149175 
\end{CodeOutput}
\end{CodeChunk}
  So here we find a median risk difference of $-0.33$ and a 95\%
  credible interval of [$-0.40$, $-0.21$] for this example. The
  risk difference distribution could now also be investigated further
  using histograms etc.

\subsubsection{Credible intervals}
  Central credible intervals can be computed using the corresponding
  posterior quantiles via the ``\code{...\$qposterior()}'' function
  (see above).  By default however, \emph{shortest} intervals (see
  Section~\ref{sec:CIs}) are provided in the \code{bayesmeta()}
  output, or they can also be computed using the
  \code{...\$post.interval()} function.  The \code{bayesmeta()}
  function's default behaviour may also be controlled by setting the
  ``\code{interval.type}'' argument.  Looking at Figure~\ref{fig:plot}
  (marginal posteriors at the bottom), one can see that, depending on
  the posterior's shape, the shortest intervals may turn out one- or
  two-sided, at least for the heterogeneity parameter. For example a
  99\% credible interval for the heterogeneity can then be computed
  via
\begin{CodeChunk}
\begin{CodeInput}
R> ma01$post.interval(tau.level=0.99)
\end{CodeInput}
\begin{CodeOutput}
[1] 0.000000 1.109186
attr(,"interval.type")
[1] "shortest"
\end{CodeOutput}
\end{CodeChunk}
  One can also see that the returned interval contains an attribute
  indicating the type of interval.  A central interval then is derived
  by explicitly specifying the method to be used for computation:
\begin{CodeChunk}
\begin{CodeInput}
R> ma01$post.interval(tau.level=0.99, method="central")
\end{CodeInput}
\begin{CodeOutput}
[1] 0.003547657 1.205400562
attr(,"interval.type")
[1] "central"
\end{CodeOutput}
\end{CodeChunk}
  Such an interval then is actually simply based on the corresponding
  ``central'' quantiles, as one may confirm by running:
\begin{CodeChunk}
\begin{CodeInput}
R> ma01$qposterior(tau.p=c(0.005, 0.995))
\end{CodeInput}
\begin{CodeOutput}
[1] 0.003547657 1.205400562
\end{CodeOutput}
\end{CodeChunk}

\subsubsection{Prediction}
  Besides inferring the ``main'' parameters $\mu$ and $\tau$, one can
  do the same computations for prediction, i.e., a future study's
  parameter~$\theta_{k+1}$.  Basic summary statistics for the
  posterior predictive distribution are already contained in the
  ``\code{...\$summary}'' element (see above).  The
  ``\code{...\$dposterior()}'', ``\code{...\$pposterior()}'',
  ``\code{...\$qposterior()}'', ``\code{...\$rposterior()}'' and
  ``\code{...\$post.interval()}'' functions all have an optional
  ``\code{predict}'' argument to request the predictive distribution.
  That way, one can for example combine the posterior and predictive densities of
  $\mu$ and $\theta_{k+1}$ in a plot:
\begin{CodeChunk}
\begin{CodeInput}
R> x <- seq(-3.5, 0.5, length=200)
R> plot(x, ma01$dposterior(mu=x), type="n",
+    xlab="effect", ylab="probability density")
R> abline(h=0, v=0, col="grey")
R> lines(x, ma01$dposterior(mu=x), col="red")
R> lines(x, ma01$dposterior(mu=x, predict=TRUE), col="blue")
\end{CodeInput}
\end{CodeChunk}
  \begin{figure}[t]
    \begin{center}
      \includegraphics[width=0.9\linewidth]{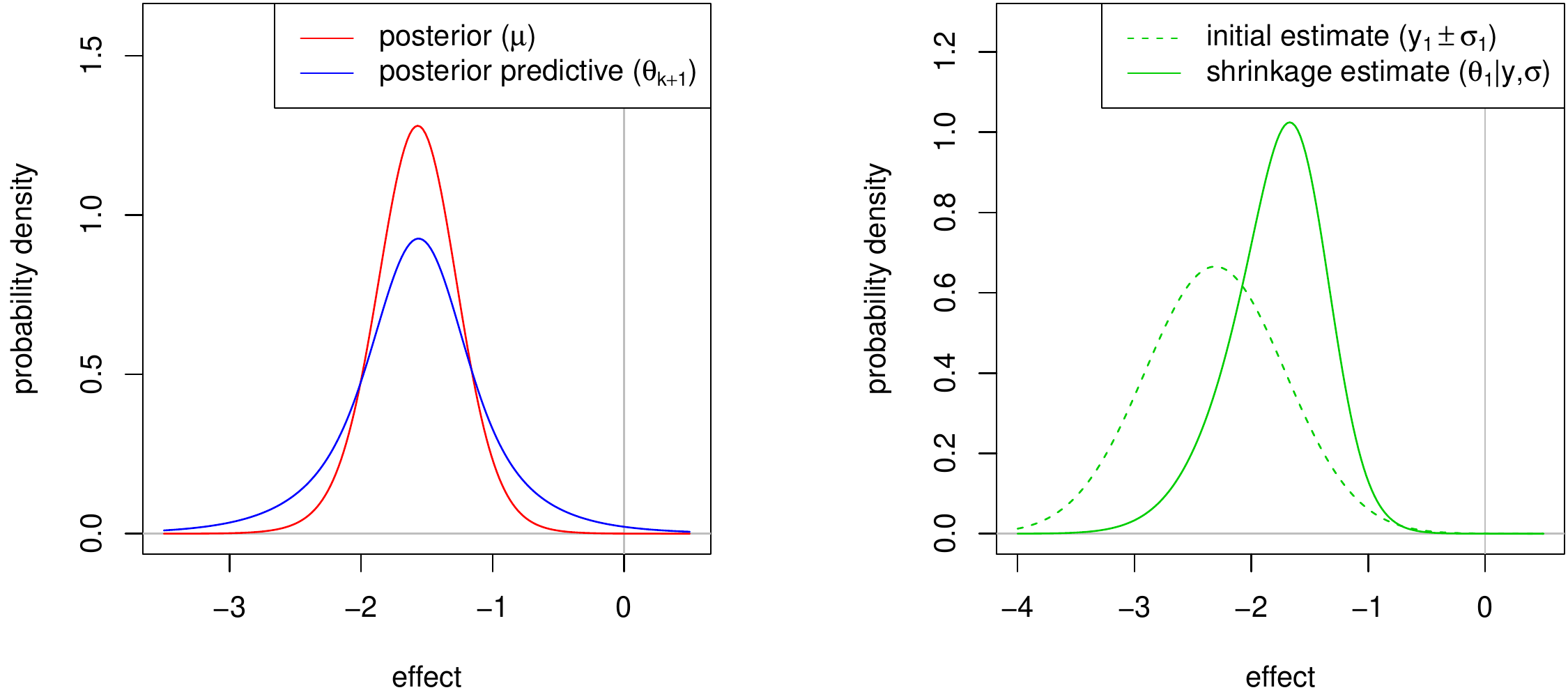}
      \caption{\label{fig:densities} Posterior and posterior
        predictive densities for the overall effect~$\mu$ and a
        `future' study's parameter~$\theta_{k+1}$ (left panel), and
        original ($y_1$, $\sigma_1$) and shrinkage
        (posterior~$\theta_1|\vec{y},\vec{\sigma}$) estimates for the
        first ($i\!=\!1$) study (right panel). The corresponding
        estimates (medians and 95\% credible intervals) are also
        shown in the forest plot in Figure~\ref{fig:forestplot01}.}
    \end{center}
  \end{figure}
  The resulting plot is shown in Figure~\ref{fig:densities} (left
  panel). Analogously, the ``\code{predict}'' argument may be used to
  compute e.g.\ CDFs, quantile functions or credible intervals.

\subsubsection{Shrinkage}
  The ``shrinkage'' posterior distributions of the study-specific
  parameters~$\theta_i$ are also accessible from the
  \code{bayesmeta()} output.  They are also summarized in the
  ``\code{...\$theta}'' element; for example, shrinkage for the first
  two studies is shown in the first two columns:
\begin{CodeChunk}
\begin{CodeInput}
R> ma01$theta[,1:2]
\end{CodeInput}
\begin{CodeOutput}
          Heffron (2003) Gibelli (2004)
y             -2.3097026     -0.4595323
sigma          0.5994763      0.5563956
mode          -1.6711220     -1.3895876
median        -1.7411356     -1.2821722
mean          -1.7778965     -1.2339736
sd             0.4229425      0.4488759
95% lower     -2.6561049     -2.0455088
95% upper     -0.9920023     -0.3089461
\end{CodeOutput}
\end{CodeChunk}
  \sloppy    
  One can see the original data ($y_i$ and $\sigma_i$) along with the
  posterior summaries (see also the forest plot in
  Figure~\ref{fig:forestplot01}).  The ``\code{...\$dposterior()}'',
  ``\code{...\$pposterior()}'', ``\code{...\$qposterior()}'',
  ``\code{...\$rposterior()}'' and ``\code{...\$post.interval()}''
  functions again also have an optional ``\code{individual}'' argument
  to specify one of the individual studies (either by their index or
  by their name). For example, one can illustrate the first study's
  ($i=1$) input data ($y_1$, $\sigma_1$) and shrinkage estimate
  ($\theta_1$) in a single plot using the following code
\begin{CodeChunk}
\begin{CodeInput}
R> x <- seq(-4, 0.5, length=200)
R> plot(x, ma01$dposterior(theta=x, individual=1), type="n",
+    xlab="effect", ylab="probability density")
R> abline(h=0, v=0, col="grey")
R> lines(x, dnorm(x, mean=ma01$y[1], sd=ma01$sigma[1]), 
+    col="green", lty="dashed")
R> lines(x, ma01$dposterior(theta=x, individual=1), col="green")
\end{CodeInput}
\end{CodeChunk}
  The resulting two densities are shown in Figure~\ref{fig:densities}
  (right panel).  Analogously, the ``\code{individual}'' argument may
  be used to compute e.g.\ CDFs, quantile functions or credible
  intervals.

\subsection{Investigating prior variations}\label{sec:priorVariations}
\subsubsection{Prior predictive distributions}
  In order to judge the implications of settings of the heterogeneity
  prior, it is often useful to consider \emph{prior predictive
    distributions} \citep{BDA3rd}. Any fixed value of~$\tau$ will
  imply a certain (prior) distribution $p(\theta_i|\mu,\tau)$ and variability among the true
  study-specific means $\theta_1,\ldots,\theta_k$, namely, a normal
  distribution with $\var(\theta_i|\mu,\tau)=\tau^2$ (see also (\ref{eqn:NNHM2})). Depending on the
  type of endpoint (e.g., log-ORs), the implied variability can be
  interpreted and judged on the corresponding outcome scale
  \citep[][Sec.~5.7]{SpiegelhalterEtAl}.

  Assuming a prior distribution for~$\tau$, rather than a fixed value,
  also implies assumptions on the a~priori expected distribution and
  variability of the true study parameters~$\theta_i$. The
  \emph{prior predictive} distribution~$p(\theta_i|\tau)$ of the $\theta_i$ values then
  is a mixture of normal distributions, with mean~$\mu$ and with the
  prior~$p(\tau)$ as the mixing distribution for the normal standard
  deviation \citep{Seidel2010,Lindsay}. As the name suggests, the
  prior predictive distribution is actually closely related to the
  (posterior) predictive distribution discussed above \citep{BDA3rd}.
  This mixture distribution can again be evaluated using the
  \textsc{direct} algorithm \citep{RoeverFriede2017}; this approach is
  implemented in the \code{normalmixture()} function.

  Consider the half-normal prior distribution with scale~$0.5$ that
  was used for the heterogeneity in the above analysis. We can now
  check what prior predictive distribution this prior corresponds
  to. We only need to supply the prior CDF (the mean~$\mu$ is by
  default set to zero):
\begin{CodeChunk}
\begin{CodeInput}
R> hn05 <- normalmixture(cdf=function(t){phalfnormal(t, scale=0.5)})
\end{CodeInput}
\end{CodeChunk}
  One can check the returned result (e.g.\ via \code{str(hn05)}); the
  result is a \code{list} with several elements, among which most
  importantly are the mixture's density, cumulative distribution and
  quantile functions (``\code{...\$density()}'', ``\code{...\$cdf()}''
  and ``\code{...\$quantile()}'', respectively).

\begin{figure}[t]
    \begin{center}
      \includegraphics[width=0.6\linewidth]{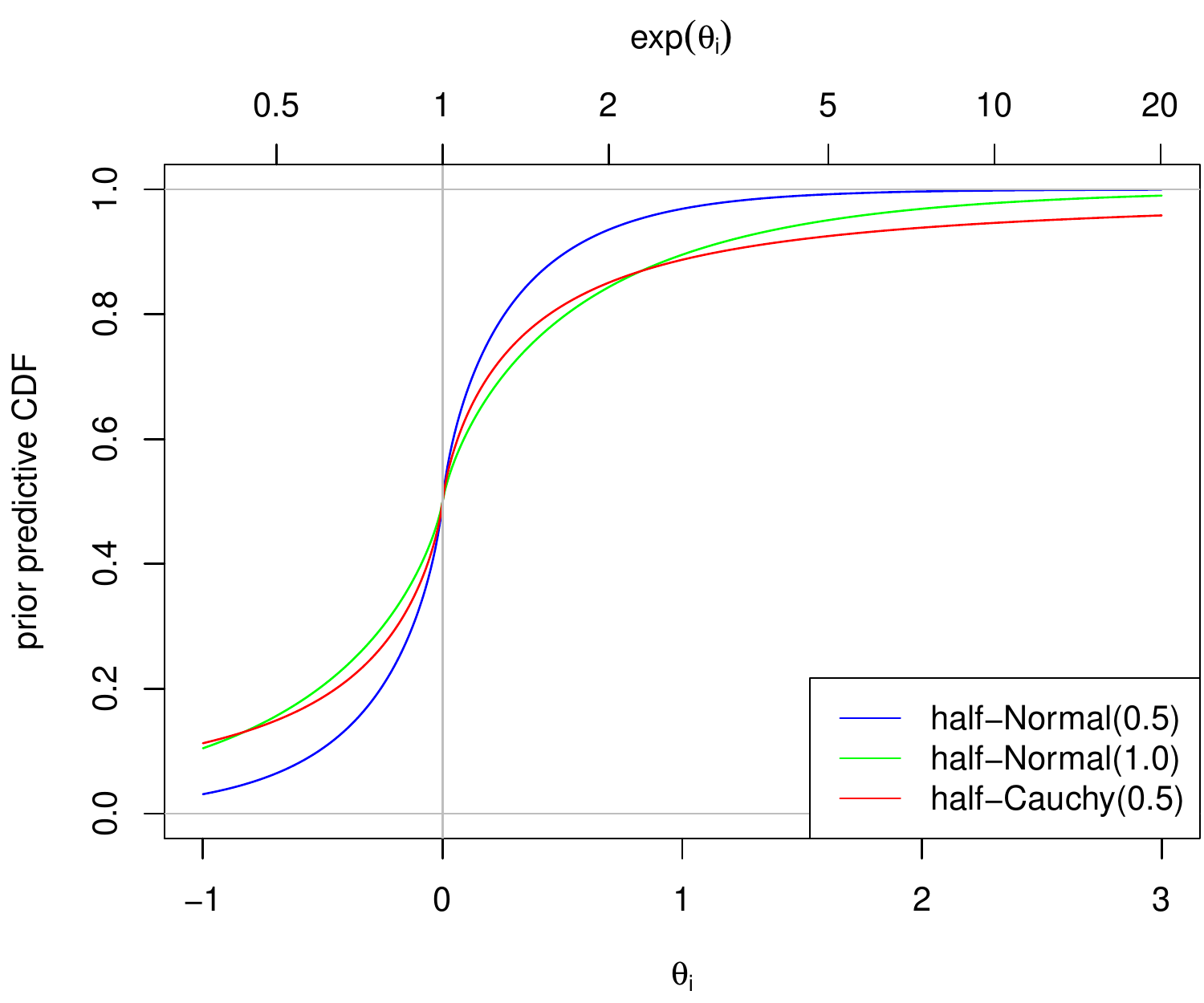}
      \caption{\label{fig:priorpredictive}Prior predictive
        distributions for the true study means~$\theta_i$ assuming
        several heterogeneity priors (and $\mu=0$) as computed using
        the \code{normalmixture()} function.}
    \end{center}
  \end{figure}

  For comparison, we can also check the implications of a half-Cauchy
  prior of the same scale, or a half-normal prior of doubled scale:
\begin{CodeChunk}
\begin{CodeInput}
R> hn10 <- normalmixture(cdf=function(t){phalfnormal(t, scale=1.0)})
R> hc05 <- normalmixture(cdf=function(t){phalfcauchy(t, scale=0.5)})
\end{CodeInput}
\end{CodeChunk}
  and compare these graphically via their implied prior predictive
  CDFs by accessing the three mixtures' ``\code{...\$cdf()}''
  functions:
\begin{CodeChunk}
\begin{CodeInput}
R> x <- seq(-1, 3, by=0.01)
R> plot(x, hn05$cdf(x), type="l", col="blue", ylim=0:1,
+    xlab=expression(theta[i]), ylab="prior predictive CDF")
R> lines(x, hn10$cdf(x), col="green")
R> lines(x, hc05$cdf(x), col="red")
R> abline(h=0:1, v=0, col="grey")
R> axis(3, at=log(c(0.5,1,2,5,10,20)), lab=c(0.5,1,2,5,10,20))
R> mtext(expression(exp(theta[i])), side=3, line=2.5)
\end{CodeInput}
\end{CodeChunk}
  The resulting plot is shown in Figure~\ref{fig:priorpredictive}.
  In our example, the effect measure is a log-OR, so the $\theta_i$
  need to be interpreted on the exponentiated scale (see the top
  axis). A~priori, $95\%$ of $\theta_i$ values are assumed to be
  within $\pm$ the $97.5\%$ quantile of the (symmetric) prior
  predictive distribution. We can now check what this means for our
  three cases:
\begin{CodeChunk}
\begin{CodeInput}
R> q975 <- c("half-normal(0.5)" = hn05$quantile(0.975),
+            "half-normal(1.0)" = hn10$quantile(0.975),
+            "half-Cauchy(0.5)" = hc05$quantile(0.975))
R> print(cbind("theta"=q975, "exp(theta)"=exp(q975)))
\end{CodeInput}
\vspace{1ex}    %%%  <---  note the hack here!
\begin{CodeOutput}
                    theta exp(theta)
half-normal(0.5) 1.092287   2.981083
half-normal(1.0) 2.184573   8.886857
half-Cauchy(0.5) 5.050571 156.111517
\end{CodeOutput}
\end{CodeChunk}
  So for the half-normal prior with scale~$0.5$ we have $95\%$
  probability roughly within a factor of~$\frac{1}{3}$ or~$3$ around
  the overall mean odds ratio~($\exp(\mu)$). For the other two priors,
  the numbers are much more extreme.

\subsubsection{Informative heterogeneity priors}
  It may also make sense to consider empirical information for the
  setup of an informative heterogeneity prior, for example, when other
  evidence is extremely sparse.  In medical or psychological contexts,
  some evidence for certain types of endpoints may be found e.g.\ in
  \citet{Pullenayegum2011}, \citet{TurnerEtAl2012},
  \citet{KontopantelisSpringateReeves2013} and \citet{vanErpEtAl2017}.
  Instantly applicable for a meta-analysis are the numbers given by
  \citet{RhodesEtAl2015} and \citet{TurnerEtAl2015}, where in both
  cases the complete \emph{Cochrane Database of Systematic Reviews}
  was analyzed to infer the predictive distribution of heterogeneity
  for specific applications. The investigation by
  \citet{RhodesEtAl2015} here was concerned with mean difference
  endpoints, while \citet{TurnerEtAl2015} focused on log-OR endpoints.
  The derived prior distributions are directly available in the
  \pkg{bayesmeta} package via the \code{RhodesEtAlPrior()} and
  \code{TurnerEtAlPrior()} functions. For our present example (a
  log-OR endpoint whose definition may be categorized as ``surgical /
  device related success / failure'', and where the comparison is
  between pharmacological treatment and control), we can derive the
  prior simply as
\begin{CodeChunk}
\begin{CodeInput}
R> tp <- TurnerEtAlPrior(outcome = "surgical", 
+    comparator1 = "pharmacological", comparator2 = "placebo / control")
\end{CodeInput}
\end{CodeChunk}
  For a complete description of the possible input options see the
  online documentation; the \code{RhodesEtAlPrior()} function then
  works similarly.  The function output is a \code{list} with several
  entries, including the prior density, cumulative distribution and
  quantile function (in this case a log-normal distribution) in the
  ``\code{...\$dprior()}'', ``\code{...\$pprior()}'' and
  ``\code{...\$qprior()}'' elements.  This way we can e.g.\ check what
  magnitude of heterogeneity values is a~priori expected for this
  setting by determining the median as well as $2.5\%$ and $97.5\%$
  quantiles:
\begin{CodeChunk}
\begin{CodeInput}
R> tp$qprior(c(0.025, 0.5, 0.975))
\end{CodeInput}
\begin{CodeOutput}
[1] 0.06233896 0.34300852 1.88734045
\end{CodeOutput}
\end{CodeChunk}
  The prior density can now immediately be used and passed on to the
  \code{bayesmeta()} function; for example, we can use the same effect
  prior as before and the ``empirical'' prior for the heterogeneity:
\begin{CodeChunk}
\begin{CodeInput}
R> ma02 <- bayesmeta(crins.es, mu.prior.mean = 0, mu.prior.sd = 4,
+    tau.prior = tp$dprior)
\end{CodeInput}
\end{CodeChunk}
  Comparing the results to the previous analysis (e.g.\ via their
  ``\code{...\$summary}'' outputs), one can see that in this case they
  are very similar. The two corresponding prior densities are also
  shown in Figure~\ref{fig:priors} (page~\pageref{fig:priors}; solid
  and dotted blue lines).

\subsubsection{Non-informative priors}
  As discussed in Section~\ref{sec:prior}, an obvious choice of an
  uninformative prior for the effect~$\mu$ would be the (improper)
  uniform prior on the real line; this one can be utilized by simply
  leaving the \code{mu.prior.mean} and \code{mu.prior.sd} parameters
  unspecified. In order to use one of the uninformative heterogeneity
  priors discussed in Section~\ref{sec:prior}, these do not need to be
  specified ``manually'' in terms of their probability density
  function; a set of priors is already pre-implemented and may be
  specified via a character string. The default setting for example is
  \code{tau.prior="uniform"}.  If one wants to use, say, the uniform
  effect prior along with the Jeffreys prior for the
  heterogeneity~$\tau$ (see Section~\ref{sec:prior} and
  \citet{BodnarEtAl2017}), one can run
\begin{CodeChunk}
\begin{CodeInput}
R> ma03 <- bayesmeta(crins.es, tau.prior="Jeffreys")
\end{CodeInput}
\end{CodeChunk}
  The complete list of possible options is described in detail in the
  online documentation.

\subsection{Making the connection with frequentist results} \label{sec:frequentistConnection}
  Frequentist and Bayesian approaches to inference within the NNHM
  framework are obviously related, and it may be interesting to
  highlight the connection between the corresponding results. A simple
  frequentist analysis may be performed e.g.\ using the \pkg{metafor}
  package's ``\code{rma()}'' function via
\begin{CodeChunk}
\begin{CodeInput}
R> ma04 <- rma(crins.es)
\end{CodeInput}
\end{CodeChunk}
  \citep{Viechtbauer2010}.  By default, the \emph{restricted ML}
  (REML) heterogeneity estimator $\hat{\tau}_{\mathrm{REML}}$ is used,
  but the exact type of estimator does not matter here. The
  heterogeneity point estimate here turns out as:
\begin{CodeChunk}
\begin{CodeInput}
R> sqrt(ma04$tau2)
\end{CodeInput}
\begin{CodeOutput}
[1] 0.4670268
\end{CodeOutput}
\end{CodeChunk}  
  and we can retrieve the effect estimate and its standard error via:
\begin{CodeChunk}
\begin{CodeInput}
R> ma04$b
\end{CodeInput}
\begin{CodeOutput}
             [,1]
intrcpt -1.591513
\end{CodeOutput}
\begin{CodeInput}
R> ma04$se
\end{CodeInput}
\begin{CodeOutput}
[1] 0.3340882
\end{CodeOutput}
\end{CodeChunk}
  In the Bayesian setup, these numbers correspond to conditional
  posterior moments of the effect
  ($\mu|\tau\!=\!\hat{\tau}_{\mathrm{REML}}$) in an analysis using the
  uniform effect prior. Such an analysis was performed in the previous
  section (uniform effect prior and Jeffreys heterogeneity prior;
  the heterogeneity prior does not matter here) and stored in the
  ``\code{ma03}'' object. From this we can retrieve the effect's
  conditional posterior moments (mean and standard deviation for
  $\tau\!=\!\hat{\tau}_{\mathrm{REML}}$) using the
  ``\code{...\$cond.moment()}'' function:
\begin{CodeChunk}
\begin{CodeInput}
R> ma03$cond.moment(tau = sqrt(ma04$tau2))
\end{CodeInput}
\begin{CodeOutput}
          mean        sd
[1,] -1.591513 0.3340882
\end{CodeOutput}
\end{CodeChunk}
  and we can see that these correspond exactly to the frequentist
  effect estimate. Both analysis approaches are related through the
  use of the same likelihood function; in the Bayesian analysis
  uncertainty (e.g.\ in the heterogeneity) is accounted for via
  integration, and a prior distribution for both parameters is
  considered.

\subsection{Posterior predictive checks}
\subsubsection{A meta-analysis of two studies}
  Posterior predictive $p$\mbox{-}values allow to quantify the consistency of
  the data with certain parametric hypotheses; see
  Section~\ref{sec:PostPred}. In the following we will determine some
  $p$\mbox{-}values from the \code{bayesmeta()} output; to this end, we will
  investigate a second meta-analysis example involving only two
  studies.

  Of the six studies considered in the pediatric transplantation
  example (see Figure~\ref{fig:forestplot01}), only two were
  randomized \citep{HeffronEtAl2003,SpadaEtAl2006}.  Since randomized
  studies are usually considered as evidence of higher quality, now
  suppose one was interested in combining the randomized studies only.
  Computations analogous to the preceding example may be done via
\begin{CodeChunk}
\begin{CodeInput}
R> ma05 <- bayesmeta(crins.es[crins.es[,"randomized"]=="yes",],
+    mu.prior.mean=0, mu.prior.sd=4, 
+    tau.prior=function(t){dhalfnormal(t,scale=0.5)})
\end{CodeInput}
\end{CodeChunk}
  \begin{figure}[t]
    \begin{center}
      \includegraphics[width=0.9\linewidth]{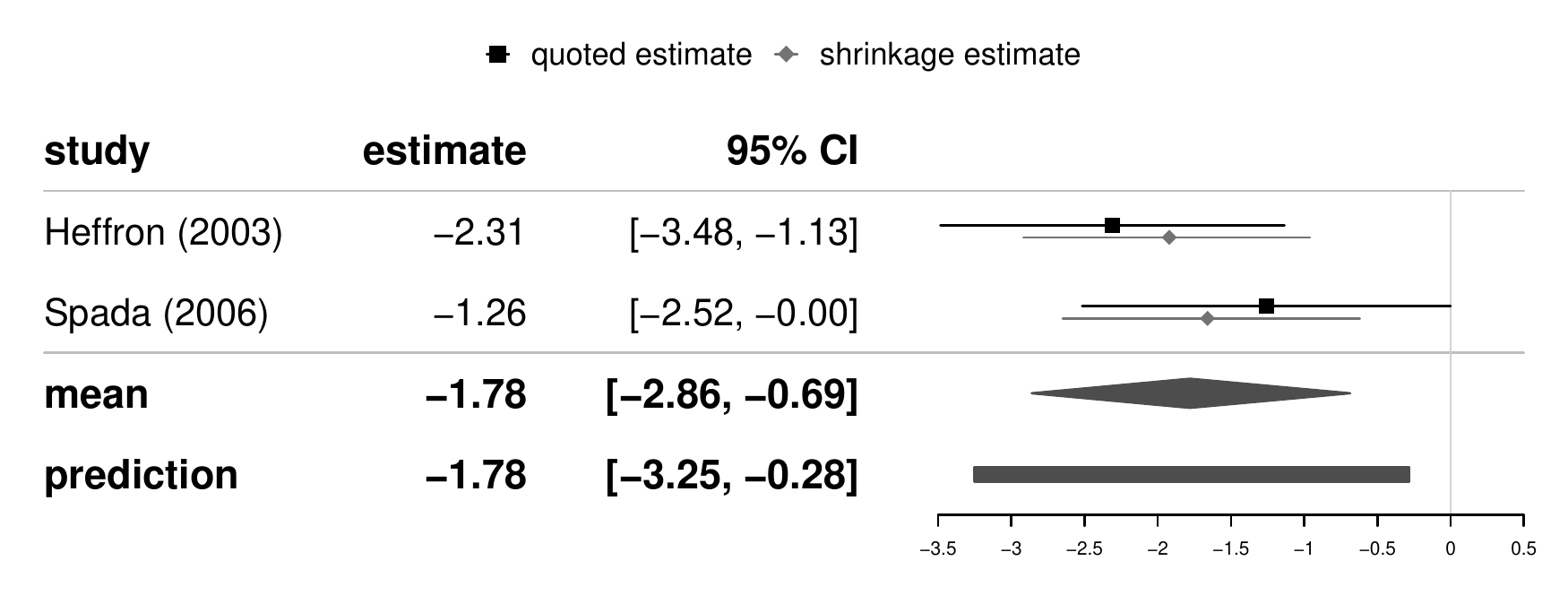}
      \caption{\label{fig:forestplot02} Forest plot 
        showing the data and derived estimates for the analysis of the two randomized studies only.}
    \end{center}
  \end{figure}
  Figure~\ref{fig:forestplot02} shows the forest plot for this
  analysis. Based on these two studies only, we can now inspect
  e.g.\ the estimate of the overall effect~$\mu$; comparing to the
  previous analysis (Figure~\ref{fig:forestplot01}), the (absolute)
  estimate is slightly larger, but the credible interval is wider.

\subsubsection[Posterior predictive p-values for the effect]{Posterior predictive $p$-values for the effect ($\mu$)}
  The obvious `null' hypothesis to be tested here is $H_0:\, \mu \geq
  0$ (i.e., no effect or a harmful effect) versus the alternative
  $H_1:\, \mu < 0$ (a beneficial effect). We may now derive a
  posterior predictive $p$\mbox{-}value in order to express to what extent
  the data are consistent with or in contradiction to the null
  hypothesis. In order to evaluate the ``discrepancy'' between data
  and null hypothesis, we need a \emph{test statistic} or
  \emph{discrepancy variable} that in some sense measures or
  captures this (in-) compatibility.
  
  An obvious candidate may e.g.\ be the posterior probability of a
  beneficial effect, $\prob(\mu < 0\,|\,y)$. This probability here is
  identical to the posterior cumulative distribution function (CDF)
  evaluated at the hypothesized value $\mu=0$.  Large values then are
  evidence \emph{against}, and small values speak \emph{in favour
    of} the null hypothesis.  In the present example data set we can
  evaluate this figure as
\begin{CodeChunk}
\begin{CodeInput}
R> ma05$pposterior(mu=0)
\end{CodeInput}
\begin{CodeOutput}
[1] 0.9974968
\end{CodeOutput}
\end{CodeChunk}
  Regarding our hypothesis setup, the question then is, how (un-)
  likely our observed value of 0.9975 is under the null hypothesis
  ($H_0:\, \mu\geq 0$). In order to answer that question, we need the
  posterior distribution of the test statistic conditional on the null
  hypothesis (and the data). Using Monte Carlo sampling, we can
  generate draws of parameters from the conditional posterior
  distribution ($\mu^\star,\tau^\star,\theta^\star \,|\, y, \mu\geq
  0$) and then generate new data based on these ($y^\star \,|\,
  \mu^\star,\tau^\star,\theta^\star$) from which we can compute
  replications of the test statistic and determine its distribution.
  
  In the \pkg{bayesmeta} package, posterior predictive checks are
  implemented in the \code{pppvalue()} function.  In order to generate
  posterior predictive draws, we need to specify the involved
  hypotheses, the test statistic, and the number of Monte Carlo
  replications to be generated; here we use $n=1000$, which may take a
  few minutes to compute:
\begin{CodeChunk}
\begin{CodeInput}
R> p1 <- pppvalue(ma05, parameter="mu", value=0, alternative="less", 
+    statistic="cdf", n=1000)
\end{CodeInput}
\end{CodeChunk}
  Since the $p$\mbox{-}value is eventually computed based on the generated
  Monte Carlo samples, a value of $n \gg 100$ will usually be
  appropriate. By default, a progress bar is shown during computation,
  allowing to estimate the remaining computation time. We can then
  inspect the result by printing the returned object:
\begin{CodeChunk}
\begin{CodeInput}
R> p1
\end{CodeInput}
\begin{CodeOutput}
	'bayesmeta' posterior predictive p-value (one-sided)

data:  ma05
cdf = 0.9975, Monte Carlo replicates = 1000, p-value = 0.01
alternative hypothesis: true effect (mu) is less than 0
\end{CodeOutput}
\end{CodeChunk}
  The default output restates the hypothesis setup and shows a
  posterior predictive $p$\mbox{-}value of 0.01. This means that in 10 of the
  1000 replications generated (1\%) the statistic was larger than our
  observed 0.9975. One can also do a quick check of the uncertainty in
  this Monte-Carlo'ed $p$\mbox{-}value using e.g.\ the \code{prop.test()}
  function, which here yields a 95\% confidence interval ranging
  roughly from 0.5\% to 2\%.

\begin{figure}[t]
    \begin{center}
      \includegraphics[width=0.9\linewidth]{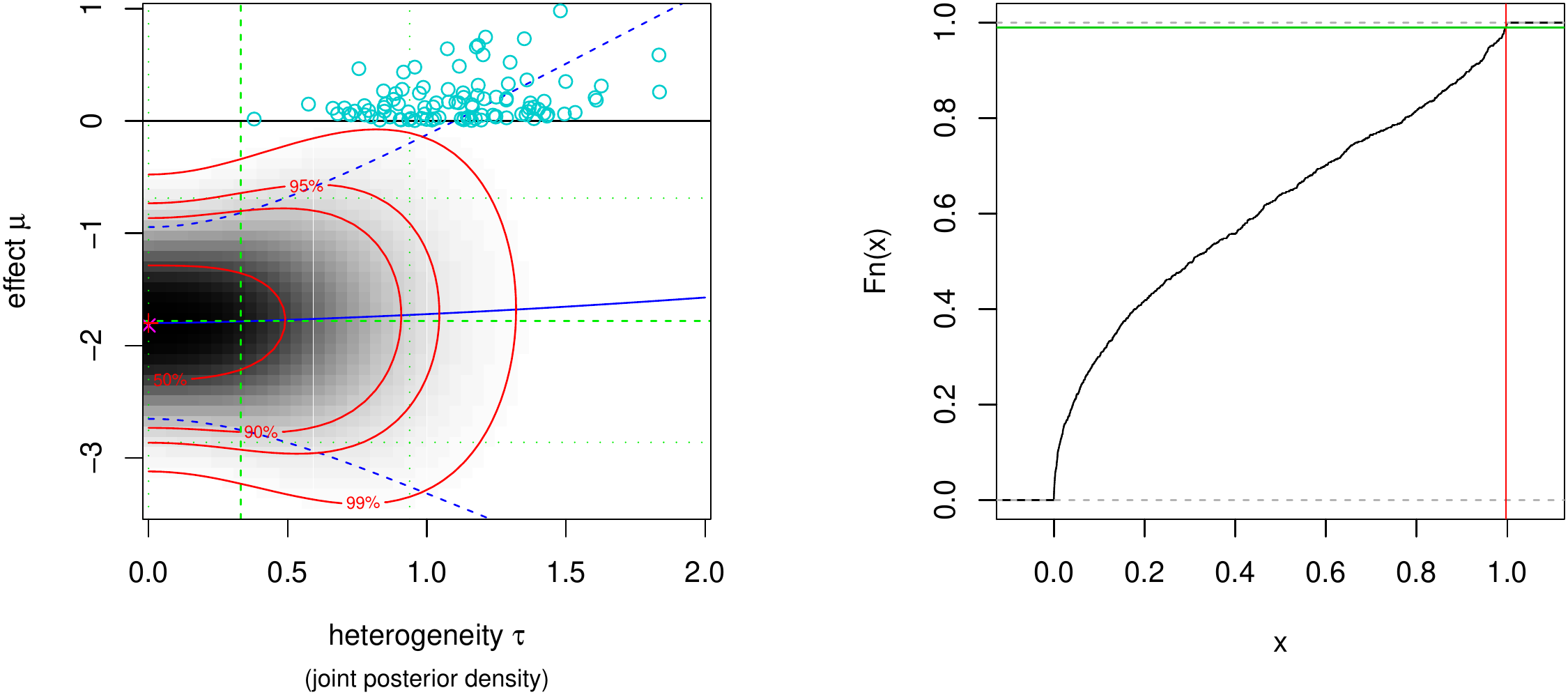}
      \caption{\label{fig:pppvalue}Illustration of the computation of
        a posterior predictive $p$\mbox{-}value using Monte Carlo
        sampling. The left panel shows the distribution of replicated
        $\mu^\star$ and $\tau^\star$ values, the right panel shows the empirical
        (cumulative) distribution of the associated ``test statistic''
        values.}
    \end{center}
  \end{figure}

  The replications are also stored in detail in the generated object
  (here: ``\code{p1}''). The \code{list} object contains a
  ``\code{...\$replicates}'' element, which again contains vectors of
  generated $\tau^\star$ and $\mu^\star$ draws, matrices of the
  corresponding $\theta^\star$ and $y^\star$ draws, and finally the
  test statistic values along with an indicator showing which ones
  constitute the ``tail area'' the $p$\mbox{-}value is based on. Using the
  provided output, one can visualize how the posterior predictive
  $p$\mbox{-}value is computed; executing
\begin{CodeChunk}
\begin{CodeInput}
R> plot(ma05, which=2, mulim=c(-3.5, 1), taulim=c(0,2))
R> abline(h=p1$null.value)                 # (the null-hypothesized mu value)
R> points(p1$replicates$tau, p1$replicates$mu, col="cyan")    # (the samples)
\end{CodeInput}
\end{CodeChunk}
  one can see the joint posterior distribution of heterogeneity~$\tau$
  and effect~$\mu$ along with the generated samples, which, according
  to the specified null hypothesis, are confined to $\mu\geq 0$ (see
  Figure~\ref{fig:pppvalue}, left panel). The resulting test statistic
  values can be illustrated via their empirical cumulative
  distribution function, which can be generated by
\begin{CodeChunk}
\begin{CodeInput}
R> plot(ecdf(p1$replicates$statistic[,1]))
R> abline(v = p1$statistic, col="red")
R> abline(h = 1-p1$p.value, col="green")
\end{CodeInput}
\end{CodeChunk}
  (see Figure~\ref{fig:pppvalue}, right panel).  The ``test
statistic'' values range between~0 and~1, and their distribution is
clearly not uniform. The actualized value in the present data set
(0.9975, vertical red line) is situated in the upper tail of the
distribution of replicated statistics values, and the remaining tail
area (horizontal green line) eventually defines the $p$\mbox{-}value.

\subsubsection[Posterior predictive p-values for the heterogeneity]{Posterior predictive $p$-values for the heterogeneity ($\tau$)}
  Computation of posterior predictive $p$\mbox{-}values for the heterogeneity
  works analogously. Use of the posterior CDF ($\prob(\tau \leq 0 |
  y)$) to test for zero heterogeneity does not make sense, as this
  figure will always be zero, for the original as well as any
  replicated data.  In order to test for zero heterogeneity, we
  could use the classical Cochran's~$Q$ statistic:
\begin{CodeChunk}
\begin{CodeInput}
R> p2 <- pppvalue(ma05, parameter="tau", value=0, alternative="greater", 
+    statistic="q", n=1000)
\end{CodeInput}
\end{CodeChunk}
  which here yields a $p$\mbox{-}value of~24.4\%. In this case computations
  are much faster, since computationally expensive re-analyses of the
  data are not necessary to compute the test statistic. The resulting
  $p$\mbox{-}value should be identical to the ``classical'' result, since
  under the null hypothesis considered (here: $\tau=0$) the
  $Q$-statistic follows a $\chi^2$-distribution, as in the frequentist
  setting.

  In a Bayesian context, it may also make sense to consider using for
  example the Bayes factor of the hypothesis of $\tau=0$ as the ``test
  statistic'' or ``discrepancy measure''. The \code{pppvalue()}
  function is able to utilize arbitrary functions as a statistic; to
  use the Bayes factor, we can define the function
\begin{CodeChunk}
\begin{CodeInput}
R> BF <- function(y, sigma) 
+  {
+    bm <- bayesmeta(y=y, sigma=sigma,
+      mu.prior.mean=0, mu.prior.sd=4,
+      tau.prior=function(t){dhalfnormal(t, scale=0.5)},
+      interval.type="central")
+    return(bm$bayesfactor[1,"tau=0"])
+  }
\end{CodeInput}
\end{CodeChunk}
  Two things are worth noting here. Firstly, it makes sense to use
  matching (especially prior) specifications for the
  \code{bayesmeta()} call within the \code{BF()} function as for the
  original analysis (here: the previously generated ``\code{ma05}''
  object). Secondly, the use of central intervals (see the
  ``\code{interval.type}'' argument) is more efficient, since these
  are faster to compute, and the intervals are otherwise irrelevant
  here. In order to utilize the function for a posterior predictive
  $p$\mbox{-}value, we can then call
\begin{CodeChunk}
\begin{CodeInput}
R> p3 <- pppvalue(ma05, parameter="tau", value=0, alternative="greater", 
+    statistic=BF, rejection.region="lower.tail", n=1000, sigma=ma05$sigma)
\end{CodeInput}
\end{CodeChunk}
  Note that the rejection region needs to be specified explicitly here
  (small Bayes factors constitute evidence \emph{against} the null
  hypothesis). Additional arguments may be passed to the
  \code{statistic} function, like the ``\code{sigma}'' argument
  above. The Bayes factor in this case yields a similar $p$\mbox{-}value to
  Cochran's $Q$ statistic ($p=22.2\%$).

\subsubsection[Posterior predictive p-values for individual effects]{Posterior predictive $p$-values for individual effects ($\theta_i$)}
  Quite commonly in a meta-analysis, interest may also be in one of
  the study specific parameters~$\theta_i$
  \citep{SchmidliEtAl2014,WandelNeuenschwanderRoeverFriede2017}. For
  example, suppose that at the end of the latter of the two concerned
  studies (Spada, 2006) a meta-analysis was performed to evaluate the
  cumulative evidence, but main interest still was in the outcome of
  the second study that had just been conducted; it would then only be
  considered in the light of the previous evidence. In such a
  scenario, we can then evaluate a posterior predictive $p$\mbox{-}value for
  the 2nd study's effect ($\theta_2$); this shrinkage estimate is also
  shown in Figure~\ref{fig:forestplot02}.  Using the \code{pppvalue()}
  function, we can simply refer to a particular study's parameter by
  its index or its label:
\begin{CodeChunk}
\begin{CodeInput}
R> p4 <- pppvalue(ma05, parameter="Spada", value=0, alternative="less",
+    statistic="cdf", n=1000)
\end{CodeInput}
\end{CodeChunk}
  which here results in a $p$\mbox{-}value of around 16.1\%.

\section{Summary}\label{sec:summary}
  A Bayesian approach has distinct advantages in the context of
  meta-analysis; it allows to coherently process the uncertainty in
  the heterogeneity (nuisance) parameter while focusing on inference
  for the effect parameter(s), small sample sizes (numbers of studies)
  do not pose a difficulty, and interpretation of the results is very
  straightforward. Since meta-analyses are quite commonly based on
  only very few studies, the opportunity to formally utilize external
  information in the analysis via the prior specification may be a
  welcome feature.  Unlike for some other methods whose results depend
  on the specification of secondary details, a Bayesian analysis
  result is uniquely defined once the model (likelihood and prior) is
  specified.

  The application of Bayesian reasoning for this purpose is not a novelty
  \citep{SpiegelhalterEtAl}, but it usually comes with a certain
  computational burden; often MCMC methods are necessary, which demand
  a substantial amount of attention on their own
  \citep{MCMCinPractice}.  The \pkg{bayesmeta} package
  \citep{bayesmeta} allows to perform Bayesian random-effects
  meta-analyses without the need to worry too much about the
  computational details. Some of the technical details of the
  computational approach underlying the package have been described
  elsewhere \citep{RoeverFriede2017}.  The simple normal-normal
  hierarchical model (NNHM) treated here is applicable in a wide range
  of contexts and is routinely used for many types of input data and
  effect measures
  \citep{HedgesOlkin,HartungKnappSinha,Viechtbauer2010,BorensteinEtAl}.
  The \pkg{bayesmeta} implementation allows for quick, accurate and reproducible
  computation, and it has already faciltated some larger-scale
  simulation studies to compare Bayesian results with common
  alternative approaches and evaluate their relative performance
  \citep{FriedeRoeverWandelNeuenschwander2017a,FriedeRoeverWandelNeuenschwander2017b}.
  Usage of the \pkg{bayesmeta} package is not more complicated to use
  than many other common meta-analysis tools.  The availability of
  predictive distributions and shrinkage estimates makes the package
  attractive also for advanced evidence synthesis applications, like
  extrapolation to future studies
  \citep{SchmidliEtAl2014,SchmidliNeuenschwanderFriede2017,WandelNeuenschwanderRoeverFriede2017}.
  Since the generic NNHM appears in different fields of application,
  use of the \pkg{bayesmeta} package may also be extended to other
  areas of research beyond common meta-analysis. For example, it could
  as well be used to model hierarchical structures \emph{within} a
  study (e.g., groups of patients), or a two-stage approach may be
  useful for meta-analysis based on individual-patient data.
  In future, the same numerical approach might be extended to the more
  general case of meta-regression.

\begin{appendix}
\section{Appendix}
\subsection{Unit information priors for binary outcomes}\label{sec:unitInfoApp}
\subsubsection{Logarithmic odds ratios (log-OR)}\label{sec:unitInfoLogOR}
  If the effect measure of an analysis is a logarithmic odds ratio
  (log-OR; see Section~\ref{sec:CrinsData}), then the standard error
  derived from a $2\!\times\!2$ contingency table amounts to
  $\sqrt{\frac{1}{a}+\frac{1}{b}+\frac{1}{c}+\frac{1}{d}}$, where $a$,
  $b$, $c$ and $d$ are the four entries (counts) and $N=a+b+c+d$ is
  the total number of subjects
  \citep{HedgesOlkin,HartungKnappSinha,BorensteinEtAl}. Assuming equal
  allocation and a neutral effect, we can simply set the table
  allocation to $a=b=c=d=\frac{N}{4}$. If we further plug in a total
  sample size of $N=1$, this leads (heuristically) to a unit
  information prior for the log-OR with zero mean and a standard
  deviation of~4.  For this prior distribution, log-ORs are
  within a range of $\pm 7.84$ with 95\% probability, corresponding to
  ORs roughly within a range from $\frac{1}{2500}$ to $2500$.

  If more generally we consider the case of a particular event
  probability~$p\in [0,1]$, we can derive a unit information prior by
  assuming $a=c=p\frac{N}{2}$ and $b=d=(1 \! - \! p)\frac{N}{2}$,
  leading to a generally even larger prior standard deviation of
  $\frac{2}{\sqrt{p(1 \! - \! p)}}$.

\subsubsection{Logarithmic relative risks (log-RR)}\label{sec:unitInfoLogRR}
  Similarly to the previous section, the logarithmic relative risk
  (log-RR) is given by $\log\Bigl(\frac{a/(a+b)}{c/(c+d)}\Bigr)$, and
  its associated standard error is
  $\sqrt{\frac{1}{a}-\frac{1}{a+b}+\frac{1}{c}-\frac{1}{c+d}}$
  \citep{HedgesOlkin,HartungKnappSinha,BorensteinEtAl}. Again plugging
  in $a=b=c=d=\frac{N}{4}$, this now amounts to a standard deviation
  of~2. If we introduce a certain event probability~$p$ (and plugging
  in $a=c=p\frac{N}{2}$ and $a+b=c+d=\frac{N}{2}$, the error is
  $2\sqrt{\frac{1-p}{p}}$, which is larger for $p<\frac{1}{2}$ and
  smaller for $p>\frac{1}{2}$.

\subsection{Conservatism of the uniform heterogeity prior}\label{sec:conservative}
  As discussed in Section~\ref{sec:prior}, it is hard to define an
  ``uninformative'' prior for the heterogeneity parameter~$\tau$.  A
  larger heterogeneity will first of all generally lead to a larger
  variance of the effect's marginal posterior (via the larger variance
  of the \emph{conditional} distribution; see
  equations~(\ref{eqn:conditionalmoments1}),
  (\ref{eqn:conditionalmoments2})).  One may then argue that an
  overestimation of heterogeneity may be considered a
  \emph{conservative} form of bias, so that, for example, among two
  exponential prior distributions the one with the larger expectation
  was ``more conservative'' in a certain sense.

  A shift in heterogeneity causes a change in \emph{both} the
  conditional standard deviation \emph{and} mean. If the shift in
  $\condmu(\tau)$ happens to be larger than the shift in
  $\condsigma(\tau)$, then the resulting (conditional) confidence
  interval for a larger heterogeneity value does not necessarily
  completely contain the interval corresponding to a smaller
  heterogeneity. Such cases may then lead to counterintuitive results
  for (frequentist) fixed- and random-effects analysis results
  especially in settings with imbalanced standard errors
  \citep{PooleGreenland1999}. Although it is not obvious whether such
  pathologies are also realistic in a Bayesian analysis,
  \emph{a~priori}, this is unlikely to lead to any systematic bias.

  Nevertheless, along these lines it is possible to show a particular
  ``conservatism'' property for the improper uniform prior. The
  derivation goes as follows.  Suppose we have a bounded parameter
  domain $[a,\infty]$, a likelihood function $f(x) \geq 0$ ($x\in
  [a,\infty]$) with $\int_a^\infty f(x) \, \differential x < \infty$,
  and a prior with monotonically decreasing probability density
  function $p(\cdot)$, so that $0 < p(a) < \infty$, and $a \leq x < y
  \;\Rightarrow\; 0\leq p(y) \leq p(x)$.  Using the (improper) uniform
  prior or prior~$p$ we get different posteriors with cumulative
  distribution functions $F_1(\cdot)$ and $F_p(\cdot)$, respectively.
  From the above assumptions follows that
  \begin{eqnarray}
    F_p(y) \;=\; \frac{\int_a^y f(x)\, p(x)\, \differential x}{\int_a^\infty f(x)\, p(x)\, \differential x}
    &\geq&
    \frac{\int_a^y f(x)\, p(y)\, \differential x}{\int_a^\infty f(x)\, p(a)\, \differential x}
    \;=\;
    \frac{p(y)}{p(a)} \,
    \frac{\int_a^y f(x)\, \differential x}{\int_a^\infty f(x)\, \differential x}
    \\
    &\geq&
    \frac{\int_a^y f(x)\, \differential x}{\int_a^\infty f(x)\, \differential x}
    \;=\; F_1(y)
  \end{eqnarray}
  for all $y>a$.  This means that with $F_p(y) \geq F_1(y)$, the
  posterior using the uniform prior is \emph{stochastically larger}
  than the posterior based on any other prior among the class of
  priors with monotonically decreasing density $p(\cdot)$ and finite
  $p(a)$ (provided the uniform prior yields a proper posterior).  In
  our context, this especially implies that quantiles or expectations
  based on the uniform prior are larger \citep{ShakedShanthikumar}.

  The class of priors with finite intercept and monotonically
  decreasing density to which the above property applies includes
  e.g.\ the exponential, half-normal, half-Student\mbox{-}$t$,
  half-Cauchy and Lomax distributions \citep{JohnsonKotzBalakrishnan},
  or uniform distributions with a finite upper bound.

\subsection{Marginal likelihood derivation}\label{sec:marginalDerivation}
  Using the improper uniform prior for~$\mu$ ($p(\mu)\propto 1$), the
  marginal likelihood, marginalized over~$\mu$, is
\begin{eqnarray}
  p(\vec{y}|\tau,\vec{\sigma}) 
  &=& 
  \int p(\vec{y}|\mu,\tau,\vec{\sigma}) \, p(\mu) \, \differential \mu
  \\
  &=&
  \bigl(2\pi\bigr)^{-\frac{k}{2}} \times
  \prod_{i=1}^k\frac{1}{\sqrt{\sigma_i^2\!+\!\tau^2}} \times
  \int \exp\Biggl(-\frac{1}{2}\sum_{i=1}^k\frac{(y_i-\mu)^2}{\sigma_i^2\!+\!\tau^2}\Biggr) \, \differential \mu \label{eqn:marginal1a}
\end{eqnarray}
where 
\begin{eqnarray}
  \sum_{i=1}^k\frac{(y_i-\mu)^2}{\sigma_i^2+\tau^2} 
  &=&
  \underbrace{\sum_{i=1}^k\frac{y_i^2}{\sigma_i^2+\tau^2}}_{=:a}
  +\mu\underbrace{\Biggl(-2\sum_{i=1}^k\frac{y_i}{\sigma_i^2+\tau^2}\Biggr)}_{=:b}
  +\mu^2\underbrace{\sum_{i=1}^k\frac{1}{\sigma_i^2+\tau^2}}_{=:c}
  \\ &=&
  a + b\,\mu +c\,\mu^2
  \\ &=&
  \frac{(\mu-\frac{-b}{2c})^2}{\sqrt{1/c}^2} + a - {\textstyle\frac{b^2}{4c}}
  \;=\;
  \frac{\bigl(\mu-\condmu(\tau)\bigr)^2}{\condsigma(\tau)^2} + \Delta(\tau) \label{eqn:conditional}
\end{eqnarray}
and 
\begin{eqnarray}
  \condmu(\tau)
  &=&
  \frac{-b}{2c} \;=\;
  \frac{\sum_{i=1}^k\frac{y_i}{\sigma_i^2+\tau^2}}{\sum_{i=1}^k\frac{1}{\sigma_i^2+\tau^2}} \label{eqn:ThetaCondMean1}
  \\
  \condsigma(\tau)
  &=&
  \sqrt{1/c} \;=\;
  \sqrt{\frac{1}{\sum_{i=1}^k\frac{1}{\sigma_i^2+\tau^2}}}\label{eqn:ThetaCondStdev1}
  \\
  \Delta(\tau)
  &=&
  a - {\textstyle\frac{b^2}{4c}} \;=\;
  \sum_{i=1}^k\frac{y_i^2}{\sigma_i^2+\tau^2}
  - \frac{\Bigl(\sum_{i=1}^k\frac{y_i}{\sigma_i^2+\tau^2}\Bigr)^2}{\sum_{i=1}^k\frac{1}{\sigma_i^2+\tau^2}}.
  \\ 
  &=&
  \sum_{i=1}^k \frac{1}{\sigma_i^2+\tau^2}\Biggl(y_i - \sum_{j=1}^k\frac{\frac{1}{\sigma_j^2+\tau^2}\;y_j}{\sum_{\ell=1}^k\frac{1}{\sigma_\ell^2+\tau^2}}\Biggr)^2
  \;=\;
  \sum_{i=1}^k \frac{\bigl(y_i - \condmu(\tau)\bigr)^2}{\sigma_i^2+\tau^2}\mbox{.}
\end{eqnarray}
  Note that $\condmu(\tau)$ (\ref{eqn:ThetaCondMean1}) and
  $\condsigma(\tau)$ (\ref{eqn:ThetaCondStdev1}) are the conditional
  posterior mean and standard deviation of $\mu|\tau$.  With that the
  marginal likelihood turns out as
\begin{eqnarray}
  p(\vec{y}|\tau,\vec{\sigma})
  &=&
  \bigl(2\pi\bigr)^{-\frac{k}{2}} \times
  \prod_{i=1}^k\frac{1}{\sqrt{\sigma_i^2+\tau^2}} \times
  \int \exp\biggl(-\frac{1}{2}\frac{\bigl(\mu-\condmu(\tau)\bigr)^2}{\condsigma^2(\tau)}-\frac{1}{2}\Delta(\tau)\biggr) \, \differential \mu
  \\
  &=&
  \bigl(2\pi\bigr)^{-\frac{k}{2}} \times
  \prod_{i=1}^k\frac{1}{\sqrt{\sigma_i^2+\tau^2}} \times
  \exp\bigl({\textstyle-\frac{1}{2}}\Delta(\tau)\bigr)
  \times
  \int \exp\biggl(-\frac{1}{2}\frac{\bigl(\mu-\condmu(\tau)\bigr)^2}{\condsigma^2(\tau)}\biggr) \, \differential \mu
  \\
  &=&
  \bigl(2\pi\bigr)^{-\frac{k}{2}} \times
  \prod_{i=1}^k\frac{1}{\sqrt{\sigma_i^2+\tau^2}} \times
  \exp\bigl({\textstyle-\frac{1}{2}}\Delta(\tau)\bigr)
  \times
  \sqrt{2\pi}\,\condsigma(\tau)
  \\
  &=&
  \bigl(2\pi\bigr)^{-\frac{k-1}{2}} \times
  \prod_{i=1}^k\frac{1}{\sqrt{\sigma_i^2+\tau^2}} \times
  \exp\biggl(-\frac{1}{2}\frac{\bigl(y_i-\condmu(\tau)\bigr)^2}{\sigma_i^2+\tau^2}\biggr)
  \times
  \frac{1}{\sqrt{\sum_{i=1}^k \frac{1}{\sigma_i^2+\tau^2}}}\mbox{.}
  \label{eqn:marginal1b}
\end{eqnarray}

  The derivation for an informative normal effect prior (with
  mean~$\priormu$ and variance~$\priorsigma^2$) works similarly.

\subsection{Mixture implementation details}\label{sec:directDetails}
  The approximation of marginal effect distributions etc.\ is
  implemented via the \textsc{direct} algorithm as described by
  \citet{RoeverFriede2017}. This approximation is utilized to evaluate
  posterior distributions of the overall effect~$\mu$ as well as
  shrinkage estimates~$\theta_i$ and predictions~$\theta_{k+1}$. In
  all three cases, the distributions of interest are mixtures of
  conditionally normal distributions; in order to construct the
  approximate discrete mixture, it is necessary to evaluate
  \emph{symmetrized divergences} of the conditional
  distributions. The symmetrized divergence (relative entropy) for two
  normal distributions with mean and variance parameters
  $(\mu_A,\sigma^2_A)$ and $(\mu_B,\sigma^2_B)$, respectively, is
  given by
  \begin{equation}
    \mathcal{D}_\mathrm{s}\bigl(p(\vartheta|\mu_A,\sigma_A)\big\|p(\vartheta|\mu_B,\sigma_B)\bigr) 
    \;=\; \textstyle
    \frac{(\mu_A-\mu_B)^2}{\left(\frac{1}{2}(\sigma_A^{-2}+\sigma_B^{-2})\right)^{-1}}
    + \frac{(\sigma_A^2-\sigma_B^2)^2}{2\,\sigma_A^2\,\sigma_B^2}
  \end{equation}
  \citep{RoeverFriede2017}.  Since the conditional means of $\mu|\tau$
  and $\theta_{k+1}|\tau$ are identical, while the conditional
  variance of the latter is always equal to or larger than the former
  (see Section~\ref{sec:shrinkage}), a grid constructed for the
  effect's posterior ($\mu$) can always be used for the predictive
  distribution ($\theta_{k+1}$) without loss of accuracy.  In order
  not to have to construct and consider several separate $\tau$~grids
  also for the different shrinkage distributions, the general
  algorithm is slightly extended. Instead of determining divergences
  corresponding to pairs of~$\tau$ values with respect to
  \emph{each} of the shrinkage distributions individually, the
  \emph{maximum} divergence across effect posterior as well as
  all~$k$ shrinkage distributions is considered. The result is a
  single grid in $\tau$~values that may be re-used for all three types
  of distributions.

\subsection{Calibration check}\label{sec:calibration}
  The inferential statements returned by a Bayesian analysis differ in
  their probabilistic claims from those returned by frequentist
  analyses. For example, while a frequentist 95\% confidence interval
  usually is supposed to yield 95\% coverage for repeated data
  generation and analysis \emph{conditional on any single point in
    parameter space}, a Bayesian analysis is to be understood
  \emph{conditional on the assumed prior distribution}, and hence the
  coverage holds for repeated sampling of parameters from the prior
  and subsequent data generation and analysis. While frequentist
  analyses often rely on large-sample-size asymptotics (here:
  large~$k$), Bayesian posterior analyses generally should (at least
  for proper priors) yield exact coverages, independent of sample
  sizes \citep{Dawid1982,GneitingEtAl2007}. The accuracy (calibration)
  of Bayesian analysis software may be checked exploiting this
  property \citep{CookGelmanRubin2006}.
  The aim of this section is to demonstrate that the \pkg{bayesmeta} 
  implementation in fact yields consistent results.

  If a Bayesian analysis method is properly calibrated, then the
  repeated subsequent generation of (i)~parameter values
  $\theta^\star$ from the prior distribution~$p(\theta)$, (ii)~data
  $y^\star$ from the conditional sampling
  distribution~$p(y|\theta^\star)$, and (iii)~posterior probabilities
  $p^\star= \prob(\theta\leq\theta^\star|y^\star)$ will yield a sample
  of so-called \emph{probability integral transform (PIT)}
  values~$p^\star$ \citep{GneitingEtAl2007}. If the implementation is
  accurate, then these PIT values follow a uniform probability
  distribution. Investigation of the empirical cumulative distribution
  function(s) of individual parameters' PIT values returns the
  empirical frequency with which a one-sided credible interval of a
  given credible level would have covered the true value across the
  generated parameter and data samples. This way it allows to
  investigate the fidelity of the analysis procedure across the
  prior's domain as well as across credible levels
  \citep{CookGelmanRubin2006}.
  For the meta-analysis problem within the NNHM, such a calibration
  check may be implemented as follows:
\begin{CodeChunk}
\begin{CodeInput}
R> mupriormean   <- 0.0
R> mupriorsd     <- 4.0
R> taupriorscale <- 0.5
R> Nsim <- 1000
R> pit <- matrix(NA, nrow=Nsim, ncol=2, dimnames=list(NULL, c("mu","tau")))
R> for (i in 1:Nsim) {
+    # generate data:
+    mu <- rnorm(n=1, mean=mupriormean, sd=mupriorsd)  # effect
+    tau <- rhalfnormal(n=1, scale=taupriorscale)      # heterogeneity
+    k <- sample(c(2,3,5,10,20), size=1)               # number of studies
+    sigma <- runif(n=k, min=0.2, max=1.0)             # standard errors
+    y <- rnorm(n=k, mean=mu, sd=sqrt(sigma^2+tau^2))  # estimates
+    # perform analysis:
+    bma <- try(bayesmeta(y=y, sigma=sigma,
+      tau.prior=function(t){dhalfnormal(t, scale=taupriorscale)},
+      mu.prior=c(mupriormean, mupriorsd)))
+    # log probability integral transform (PIT) values:
+    if (!is.element("try-error", class(bma))) {
+      pit[i,"mu"]  <- bma$pposterior(mu=mu)
+      pit[i,"tau"] <- bma$pposterior(tau=tau)
+    }
+  }
\end{CodeInput}
\end{CodeChunk}
  Prior parameters are set at the beginning, and matching settings are
  used for the analysis. Sample sizes ($k$) here are varied between~2
  and~20, and standard errors ($\sigma_i$) between 0.2 and 1.0\@.  The
  ``\code{pit}'' matrix consists of two column vectors of PIT values
  for the marginal effect ($\mu$) and heterogeneity ($\tau$)
  posteriors, respectively. We may now illustrate the empirical
  cumulative distribution function of the $1000$~PIT values
  e.g.\ using the following commands:
\begin{CodeChunk}
\begin{CodeInput}
R> plot(ecdf(pit[,"mu"]), col="blue", 
+    main="effect (mu)", xlab="PIT value", ylab="CDF")
R> lines(0:1, 0:1, lty="dashed", lwd=2)
R> plot(ecdf(pit[,"tau"]), col="red", 
+    main="heterogeneity (tau)", xlab="PIT value", ylab="CDF")
R> lines(0:1, 0:1, lty="dashed", lwd=2)
\end{CodeInput}
\end{CodeChunk}
  \begin{figure}[t]
    \begin{center}
      \includegraphics[width=0.9\linewidth]{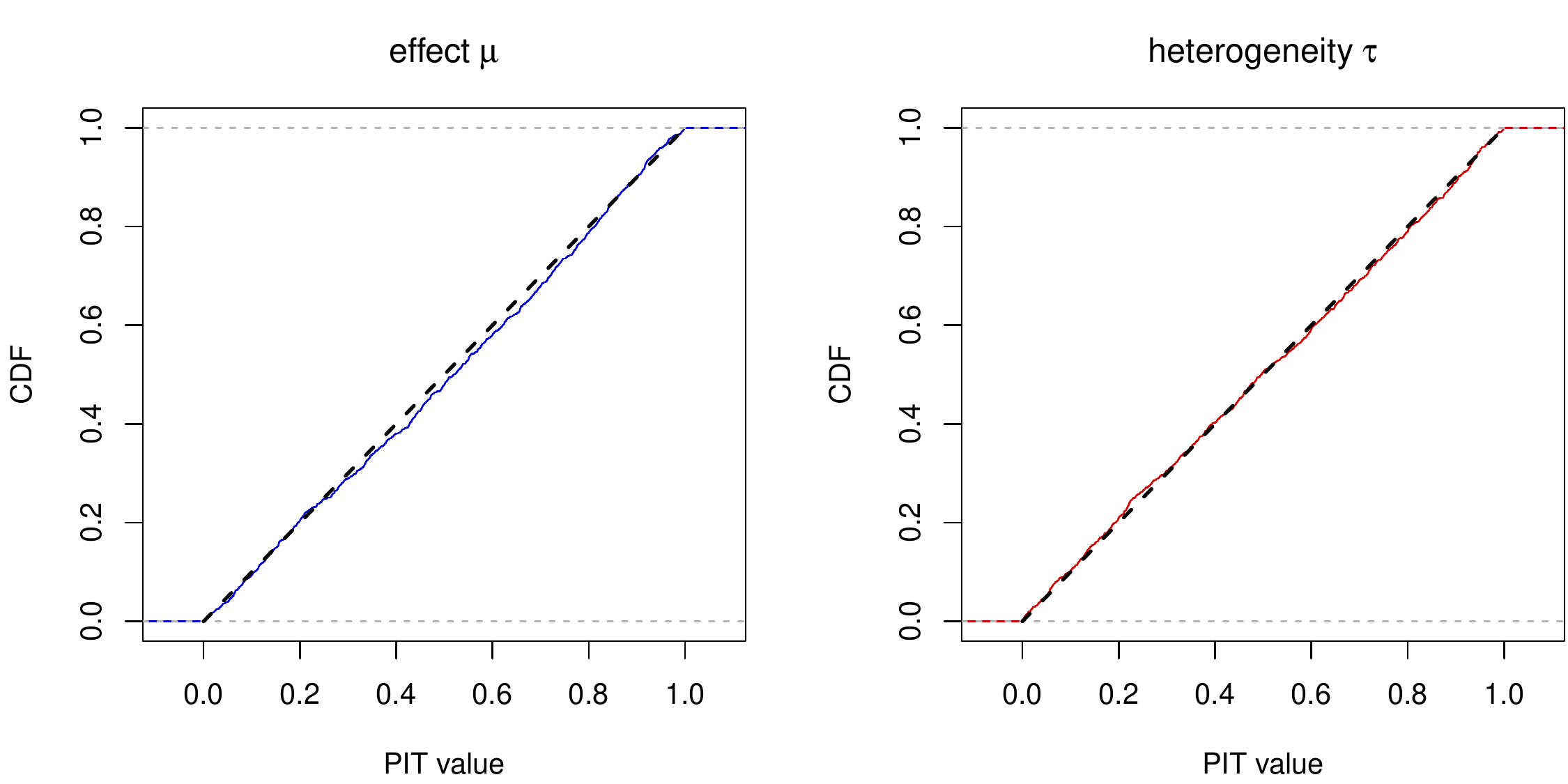}
      \caption{\label{fig:calibration}Empirical cumulative
        distribution functions of a sample of PIT values for effect
        and heterogeneity parameters. If the analysis is properly
        calibrated, the PIT values should follow a uniform
        distribution (black dashed line).}
    \end{center}
  \end{figure}
  (see Fig.~\ref{fig:calibration}). The black dashed lines here
  indicate the limiting uniform distribution that should be approached
  for large numbers of simulations. The empirical distribution is in
  close agreement with the uniform distribution here. 

  What one can read off from the plots directly is the empirical
  coverage of one-sided upper credible limits. For example, one-sided
  95\% credible limits empirically exhibited a coverage of close to
  95\% in the simulations. For the heterogeneity, a curve above the
  main diagonal may be interpreted as ``conservative'' (in the sense
  of a tendency to overestimate heterogeneity), while for the effect,
  a conservative procedure should yield a curve below the diagonal at
  the lower end and above the diagonal at the upper end (i.e., leading
  to intervals that tend to be wider than necessary).

\end{appendix}

\section*{Acknowledgements}
  This project has received funding from the European Union's Seventh
  Framework Programme for research, technological development and
  demonstration under grant agreement number FP~HEALTH~2013-602144
  through the \emph{InSPiRe} project.
  Much of the present work evolved in close collaboration with Tim
  Friede, Beat Neuenschwander and Simon Wandel. Many thanks for
  helpful comments go to Thomas Asendorf, Burak G\"{u}nhan, Markus
  Harden, Judith Heinz, Barbora Kessel, Tobias M\"{u}tze, Sibylle
  Sturtz, Steffen Unkel, and an anonymous reviewer.

{
  \bibliography{/home/christian/literature/literature}
}

\end{document}